\documentclass[a4paper,11pt]{article}
\pdfoutput=1 % if your are submitting a pdflatex (i.e. if you have
\usepackage{jheppub} % for details on the use of the package, please
\usepackage[T1]{fontenc} % if needed
\usepackage{graphicx}
\usepackage{placeins}
\usepackage{xspace}
\usepackage{soul}
\usepackage{siunitx}
\usepackage{upgreek}
\usepackage{subcaption}
\usepackage{comment}
\usepackage[dvipsnames]{xcolor}

\def\be{\begin{equation}}
\def\ee{\end{equation}}
\def\bea{\begin{eqnarray}}
\def\eea{\end{eqnarray}}

\def\iptd{\langle IP_{\mathrm{2D}} \rangle }
\def\atd{\alpha_{\mathrm 3D}}
\def\pudz{PU_{\mathrm{dz}}}
\def\mx{m_{\mathrm X}}
\def\mpi{m_{\pi_{\mathrm D}}}
\def\ctau{c \tau_{\pi_{\mathrm D}}}

\preprint{\begin{flushright} IFIC 23-24 
\end{flushright}}
\title{\boldmath Emerging jet probes of strongly interacting dark sectors}

\author[a]{Juliana Carrasco,}
\author[a]{Jose Zurita}

\affiliation[a]{Instituto de F\'isica Corpuscular, CSIC-Universitat de Val\`encia, \\ Catedr\'{a}tico Jos\'{e} Beltr\'{a}n 2, E-46980, Paterna, Spain}

\emailAdd{Juliana.Carrasco@ific.uv.es}
\emailAdd{Jose.Zurita@ific.uv.es}

\abstract{
A strongly interacting dark sector can give rise to a class of signatures dubbed dark showers, where in analogy to the strong sector in the Standard Model, the dark sector undergoes its own showering and hadronization, before decaying into Standard Model final states. When the typical decay lengths of the dark sector mesons are larger than a few centimeters (and no larger than a few meters) they give rise to the striking signature of emerging jets, characterized by a large multiplicity of displaced vertices.

In this article we consider the general reinterpretation of the CMS search for emerging jets plus prompt jets into arbitrary new physics scenarios giving rise to emerging jets. More concretely, we consider the cases where the SM Higgs mediates between the dark sector and the SM, for several benchmark decay scenarios. Our procedure is validated employing the same model than the CMS emerging jet search. We find that emerging jets can be the leading probe in regions of parameter space, in particular when considering the so-called \emph{gluon portal} and \emph{dark photon portal} decay benchmarks. With the current 16.1 fb$^{-1}$ of luminosity this search can exclude down to ${\cal O} (20) \% $ exotic branching ratio of the SM Higgs, but a naive extrapolation to the 139 fb$^{-1}$ luminosity employed in the current model-independent, indirect bound of 16 \% would probe exotic branching ratios into dark quarks down to below 10 \%. Further extrapolating these results to the HL-LHC, we find that one can pin down exotic branching ratio values of 1\%, which is below the HL-LHC expectations of 2.5$-$4 \%. We make our recasting code publicly available, as part of the LLP Recasting Repository.
} 

\begin{document} 
\maketitle
\flushbottom

\section{Introduction}
\label{s.intro}
Extensions of the Standard Model with a strongly coupled, non-Abelian dark sector have received considerable attention in the recent years. The phenomenological possibilities are varied, due the large number of parameters, such as the gauge group dimension (number of dark colors), the matter field content (number of dark flavors), the mass hierarchies in the dark sector, and the coupling strengths between the dark sector and the Standard model, as well as the internal dark sector couplings. Considering a collider operating at a center of mass energy $\sqrt{s}$, the subclass of models which confine at a scale $\Lambda_D$, and where the dark sector masses $m_D \lesssim \Lambda_D \ll \sqrt{s}$ give rise to a class of signatures generically dubbed \emph{dark showers}, in analogy to the familiar parton shower in the strong sector of the Standard Model, see~\cite{Albouy:2022cin} for a comprehensive review of the current phenomenological, experimental and theoretical status. Dark showers give rise to uncommon, exotic collider objects, such as trackless jets~\cite{Daci:2015hca}, emerging jets~\cite{Schwaller:2015gea}, semi-visible jets~\cite{Cohen:2015toa}, dark jets~\cite{Park:2017rfb} and SUEPs~\cite{Knapen:2016hky}. An experimental program aiming at dark shower signatures at the LHC is already underway~\cite{CMS:2018bvr,CMS:2021rwb,CMS:2021dzg,ATLAS:2023swa}.

It is customary~\cite{Alimena:2019zri} to dissect the collider phenomenology of dark showers into three pieces: i) \emph{production}, ii) \emph{showering} (which actually includes hadronization) and iii) \emph{decay}. The production part consists of the parton-level production of dark quarks (customarily through $2 \to 2$ processes), the showering phase includes both the emission of dark gluons and dark quarks and the formation of bound states (dark hadrons), akin to the known behaviour of the strong sector of the Standard model. Indeed, the showering process yields a distinctive signal, as, unlike in normal searches for new phenomena, one is not targeting one or two new particles, but in principle many of them. Hence, the large multiplicity inherent to the dark parton showering is what makes them stand out from other Beyond Standard Model (BSM) searches. The decay modes of these dark hadrons (including SM particles such as leptons, quarks, gluons, or potentially not decaying at all contributing to the overall dark matter abundance) combined with the lifetime spectra of these dark hadrons span a large number of phenomenologically distinct scenarios. Decay benchmarks to guide the experimental exploration have been recently put forward in~\cite{Knapen:2021eip,Knapen:2022afb}.

If the dark sector contains some particles that are long-lived (another research direction that has received considerable attention in the recent past, see e.g.~\cite{Alimena:2019zri,Lee:2018pag,Knapen:2022afb} for reviews), with lifetimes in the few mm to few meters, which decay into SM particles, they give rise to \emph{emerging jets} (EJ).  In this work we present a simple and flexible reinterpretation of the CMS search for emerging jets~\cite{CMS:2018bvr} seizing all publicly available information. We validate our procedure by reproducing the CMS results for their benchmark model, proposed in~\cite{Schwaller:2015gea}. The software developed for the reinterpretation procedure has been uploaded to the LLP Recasting Repository~\cite{LLPrepo}, which we expect to be useful for those interested in a straightforward reinterpretation of this study. We later apply our procedure to the concrete case of exotic Higgs decays, namely we obtain bounds on the branching ratio of the SM-like 125 GeV Higgs boson into a pair of dark quarks. We also show that the bounds arising from the reinterpretation of the emerging jet search, albeit not designed to target this scenario, can nonetheless provide leading constraints in large portions of the parameter space.

This article is organized as follows. In section~\ref{s.model} we briefly review the phenomenology of emerging jets and the t-channel models from~\cite{Schwaller:2015gea} which give rise to them. In section~\ref{s.ejcms}, we show our validation of the CMS emerging jet search. In section~\ref{s.results} we apply our recasting procedure to a series of BSM decay benchmarks devised in ~\cite{Knapen:2021eip}. We reserve section~\ref{s.conclu} for conclusions. Technical details about the validation of the CMS Emerging Jet is described in Appendix~\ref{a.CMS}.

\section{Emerging jet phenomenology}
\label{s.model}
When considering a strongly interacting dark sector, the known behaviour of QCD provides a guidance for the relevant phenomenological features of the model. A QCD-like sector with gauge group $SU(N_{C_D})$ and $N_{f_D}$ degenerate Dirac fermions (dark quarks, $q_{D}$) would then exhibit asymptotic freedom if $N_{f_D} < 4 N_{C_D}$, and confine at a scale $\Lambda_{D}$, where $m_{q_D} \lesssim \Lambda_{D}$. When these particles are \emph{produced} at a collider with a center-of-mass energy $\sqrt{s} < \Lambda_D$\footnote{When $\Lambda_{D}$ is close to the available energy $\sqrt{s}$, then the dark quarks have limited phase space to shower, leading to lower dark hadron multiplicities. The formed dark mesons can then be probed by searches for resonant bound states  see e.g~\cite{Kribs:2018ilo,Hochberg:2015vrg,Butterworth:2021jto}.}, the dark quarks \emph{hadronize} into dark hadrons ($\pi_D$, $\rho_D$, $\omega_D$, ...) which are clustered into collimated dark jets. The resulting signatures will then depend on the dark hadron \emph{decay}, more specifically on the lifetime spectrum $c \tau_D$, but the main underlying theme is that the shower process can lead to a large multiplicity of objects, while in traditional searches only one or two new particles are targeted.

Following~\cite{Alimena:2019zri}, the dark shower can be decomposed into the three parts described above: production, hadronization (in the dark sector) and dark hadron decays. For the production at a collider, is necessary to connect the Standard Model to the dark sector, which is done through a \emph{portal}. For emerging jets, the focus of this work, the model proposed in~\cite{Bai:2013xga,Schwaller:2015gea} employs a 
a bi-fundamental scalar $X$ (namely, charged under both QCD and the dark sector) which interacts with a SM down-type quark and a dark quark via the following Lagrangian
\be
\label{eq:EJlag}
{\cal L} \supset - \kappa_{ij} \bar{q}_{R\, i} q_{D_j} X \, \, .
\ee
While in principle $\kappa$ is a $3 \times N_{C_D}$ matrix, we consider here the case where there is one single universal coupling to the right-handed down type quark, to avoid bounds from flavour physics (FCNCs, neutral meson-mixing, rare decays). In this model, $X$ is pair produced with a sizable rate through gauge interactions, and the subsequent $X \to d_R~q_D$ decay (which happens with 100 \% branching ratio) leads us to expect two light jets and two emerging jets. Since the mediator particle $X$ appears in t-channel exchanges, the model of equation~\ref{eq:EJlag} is colloquially referred to as a \emph{t-channel} model. 

In this paper we will also study the production via an s-channel SM Higgs boson $h$, which falls in the category of exotic Higgs decays~\cite{Curtin:2013fra,Cepeda:2021rql}. Measurements of the Standard Model Higgs properties set Br($h \to {\rm exotic} < 0.16)$ at 95 \% C.L. from both ATLAS~\cite{ATLAS-CONF-2021-053} and CMS~\cite{CMS:2022dwd} with a total integrated  luminosity of 139 and 138 fb$^{-1}$ respectively. These bounds are not completely model independent, as they assume that the Higgs boson couples to the electroweak gauge bosons with a strength equal or less to the SM one, which can be violated in certain BSM scenarios. The combination of ATLAS and CMS with 3000 fb$^{-1}$ is expected to set a limit of 2.5\%, under the assumption that the current systematic uncertainties would be halved~\cite{Cepeda:2019klc}, or 4\% with the current systematic uncertainties~\cite{deBlas:2019rxi}. This leaves ample room for the $h \to q_D q_D$ to occur with sizable rates. We note that the identification of a resonant, long-lived $q_D q_D$ topology using Machine Learning techniques has been recently studied in reference~\cite{Lu:2023gjk} for the SM Higgs and in reference~\cite{Bardhan:2023mia} for a TeV scale Z'.

The showering and hadronization in the dark sector is conducted through the Hidden Valley module~\cite{Carloni:2010tw,Carloni:2011kk} within Pythia8~\cite{Sjostrand:2014zea}. The non-perturbative nature of the dark QCD-like theories prevents from consistently connecting UV and IR parameters based on a perturbative approach, and hence it is customary to consider a dark sector consisting of spin-1 dark vector mesons $\rho_D$, and of spin-0 pseudoscalar dark pions, $\pi_D$. As the $\pi_D$ arise from a breaking of a chiral dark symmetry, they are parametrically lighter than the other mesons in the theory, which decay into dark pions if kinematically allowed. Hence, the phenomenology is dictated by the dark pion properties, in particular their lifetime $c \tau_{\pi_D}$. We distinguish three possible regimes depending on the dark pion lifetime. If the dark pions decay promptly ($c \tau_{\pi_D} \lesssim 1$ mm), they end up giving multi-jet signals\footnote{If some dark hadrons would remain stable, one would have an invisible component within a jet, giving rise to semi-visible jets~\cite{Cohen:2015toa}. }. If, on the contrary, the dark pions are stable in the detector ($c \tau_{\pi_D} \gtrsim 1$ m) then they appear as missing energy, and can be targeted by the suite of missing energy signatures that are customarily searches in the dark matter program at the LHC~\cite{Boveia:2018yeb}. For $c \tau_{\pi_D} \in [0.001 - 1 ]$ m, the dark pions decay inside the detector volume with different decay lengths, depending on their boost and on the fact that their actual decay position is sampled from an exponential distribution.

The decay patterns of the dark pions can be quite varied. As mentioned before, in reference~\cite{Schwaller:2015gea} to avoid dealing with non-trivial bounds from flavour processes, a 100 \% decay rate to right-handed down quarks was assumed. Yet, the possibilities for the decay are quite numerous, and in that light reference~\cite{Knapen:2021eip} proposed five decay benchmark models, dubbed \emph{decay portals}, based on a minimal set of theoretical priors. These decay portals describe how the pseudoscalar and vector dark mesons decay into Standard Model particles~\footnote{In reference~\cite{Knapen:2021eip} the pseudoscalar and vector mesons are noted as $\tilde{\eta}$  and $\tilde{\omega}$. For a unified notation in this paper, we replace them by $\pi_D$ and $\rho_D$, respectively. }. If $\pi_D$ decays into gluons (photons) through a dimension 5 operator we have the gluon (photon) portal. If $\pi_D$ instead couples to the Standard Model Higgs via the $H^{\dagger} H$ operator, then we have the Higgs portal, where the decays to the SM quarks follow the Yukawa hierarchy of a SM-like Higgs with $m_H = m_{\pi_D}$. The other two decay portals have the $\pi_D$ decaying either through its mixing with the photon (akin to the $\gamma$-$\rho$ mixing in the Standard Model), or through the chiral anomaly into a pair of dark photons $A^{'}$, inspired by the $\pi_0 \to \gamma \gamma$ SM process. The corresponding Pythia configuration cards for each of these portals can be generated through the public python script~\cite{DStool}.

The number of free parameters in both our scenarios is still quite large, and here we follow some additional choices made in the literature. Regarding the t-channel EJ model, $N_{C_D}=3$ and $N_{f_D}=7$ are inspired by the study of~\cite{Bai:2013iqa}, and the dark sector mass parameters are chosen in proportion $\Lambda_D: m_{q_D}: m_{\rho_D}: \mpi$ being $2:2:4:1$. This choice ensures that the vector meson always decays into two dark pions, and was followed by the CMS collaboration in their emerging jets search~\cite{CMS:2018bvr}~\footnote{The inclusion of a non-trivial flavour structure, where $\Lambda_D > m_{q_D}$ is assumed, was studied in~\cite{Renner:2018fhh}.}. 
The free parameter for the analysis are then $\mx$, $\mpi$ and $\ctau$. Regarding the s-channel SM Higgs production with its different decay portals we assume $N_{C_D}=3$ and $N_{f_D}=1$, and $\Lambda_D: m_{q_D}: m_{\rho_D}: m_{\pi_D}$ is now $2.5:0.4:2.5:1$. Since the mediator mass is known, the only free parameters of the model are $\mpi$, $\ctau$ and the exotic branching ratio, $H \to q_D q_D$. We note that while in reference~\cite{Knapen:2021eip} a minimum proper lifetime as a function of the mass for each decay portal was estimated from theoretical considerations, we prefer to remove those prejudices and consider the three parameters as fully independent and uncorrelated. 

A few details about our simulations are in order. Within the Pythia Hidden Valley module, we set the parameter \texttt{HiddenValley:pTminFSR} to 1.1 $\Lambda$, and the \texttt{probVector} flag is set to 0.318, following the considerations discussed in Appendix A of~\cite{Knapen:2021eip}. In section~\ref{s.ejcms} we employ Pythia version 8.212, used in the CMS study,  as our aim is to reproduce the published limits. For section~\ref{s.results} we employ instead Pythia 8.307, because (as explained in~\cite{Albouy:2022cin}) this version has corrected a previous flaw in the code, that tended to overproduce hidden hadrons at very low $p_T$.

\section{Validation of the CMS emerging jet search}
\label{s.ejcms}
In this section we discuss in detail the validation of the CMS search for emerging jets using a total integrated luminosity of 16.1 fb$^{-1}$~\cite{CMS:2018bvr}. 

The CMS collaboration targets the dark QCD model from Equation~\ref{eq:EJlag}, via $p p \to X X$ followed by  $X \to q ~q_D$. Hence naively one expect to find two emerging jets and two SM jets. Events are selected by passing the $H_T > 900 $ GeV trigger, where $H_T$ is the scalar sum of the transverse momenta of all hadronic jets in the event, clustered with $R=0.4$ using the anti-$k_T$ algorithm~\cite{Cacciari:2008gp} applied to all tracks with $p_T > 1$ GeV. These events are required to have at least four jets within $|\eta| < 2.0$, and they undergo a further selection using kinematical variables to tag these jets as ``emerging'' and to define signal regions (called sets in the CMS paper). The explicit requirements are collected in Appendix~\ref{a.CMS}, together with the 95 \% C.L. limit to the number of signal events in each selection set, $S_{95}^{i}$.

The total number of signal events in each set can be computed as
\be
N_S^{i} (\mx, \ctau, \mpi) = \sigma (p p \to X X) \times ({\rm BR(X \to q~q_D))^2} ~\times A_i (\mx, \ctau, \mpi) \times {\cal L}
\label{eq:n95}
\ee
where ${\cal L}$ is the total integrated luminosity, $A_i$ is the acceptance for the $i-$th set number~\footnote{In this article, we will use the terms ``efficiency'' and ``acceptance'' interchangeably to denote the $A_i$ function.} and the production rate $\sigma ( p p \to X X \to q q_D q q_D)$ has been decomposed into the pair production cross section for $X$ pairs (which proceeds through gauge couplings and hence is independent of $\mpi$ and $\ctau$) times a branching ratio of $X \to q q_D$, which we set to unity along this work~\footnote{We have explicitly verified that the narrow width approximation is fulfilled in our model points, which allows us to factorize the total rate into a cross section and a branching ratio.}. To benchmark the search, CMS considered the following parameters: 
\begin{itemize}
    \item $\mx$ [GeV]=\{400,600,800,1000,1250,1500,2000\} .
    \item $\mpi$ [GeV]=\{1,2,5,10\} .
    \item $\ctau$ [mm]=\{1,2,5,25,45,60,100,150,225,300,500,1000\}.
\end{itemize}
For the $\mpi=5 $ GeV case, CMS has provided the acceptances $A_i$ in the $\ctau - \mx$ plane, indicating which of the seven selection numbers is the most sensitive one. In other words, for each $\ctau - \mx$ point scanned, with $\mpi=5$ GeV, only one of the possible seven $A_i$ functions is given.

To validate the search, we proceed in three steps. First, we check that using the provided $A_i$ efficiencies we can reproduce the published 95 \% C.L exclusion limits. In a second step, we check the degree of accuracy that we obtain for the two published kinematic distributions for the emerging jet tagging variables, and finally we show that we can reproduce with reasonable accuracy the said efficiencies and exclusion limits. The last step is crucial for our analysis, since this is what allows for reinterpretation, i.e.: derive limits from a experimental search on a model that has not been targeted by the experimental collaboration.

In what follows, we define ``exclusion'' by requiring that the ratio of our predicted number of events from equation~\ref{eq:n95} over the excluded one, 
\be
R_{\mathrm{95}} = \max_{i} \Biggl\{ \frac{N_S^{i}}{S_{95}^{i}}  \Biggr\} \, ,
\ee
is equal to unity, which is a common practice when performing reinterpretations~\cite{LHCReinterpretationForum:2020xtr}.

\subsection{Exclusion using published efficiencies}
\label{ss.pubAi}

\begin{figure}
    \centering
    \includegraphics[width=0.55\textwidth]{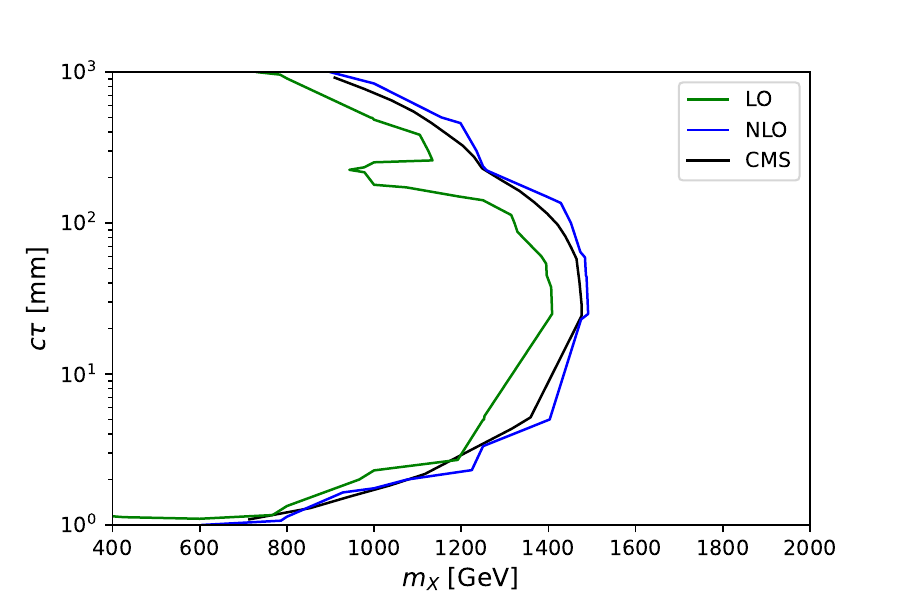}
    \caption{
    Published signal exclusions from CMS (solid black) and those obtained using the CMS acceptances and for $p p \to X X$ i)  leading order cross section in green  and ii) NLO cross section in blue.
  }
    \label{fig:limits}
\end{figure}

We start by comparing the published exclusion limit with those that can be derived from using the $A_i$ map from~\cite{CMS:2018bvr}. We present our results in figure~\ref{fig:limits}, where the published CMS exclusion is shown as solid black. For our results, we need to provide a production cross section for the $p p \to X X $ process. On one hand, we use the cross section reported during the run of Pythia 8, which is a leading-order (LO) result in the strong QCD coupling $\alpha_S$, shown in green. Second, we employ the cross sections used by CMS from~\cite{NLOcms} corresponding to down-type squark pair production, computed at the next-to-leading order in perturbation theory, and including a next-to-leading logarithmic correction from soft gluon resummation~\cite{Borschensky:2014cia}, which is displayed in blue. 
We conclude that the provided $A_i$ values are self-consistent, and we also verify that the limits were derived using NLO cross sections. 

\subsection{Kinematic distributions}
\label{ss.kin}
As a second step of our validation, we will employ the published kinematic distributions. To tag the jets passing the selection as \emph{emerging}, the following track-based variables~\cite{CMS:2017poa} are considered:
\begin{itemize}
    \item $\iptd$: the median of the unsigned transverse impact parameter. 
    \item $\pudz$: distance between the $z$ position of the primary vertex (PV), $z_{ \mathrm{PV}}$ and the $z$ position of the track at its closest approach to the PV.
    \item $D_{\mathrm N}$: the 3-D distance between the track and the primary vertex, weighted by the inverse resolution, \begin{equation}  D_\mathrm{N}^2 = \Bigl ( \frac{z_{\mathrm{PV}} - z_{\mathrm{trk}}} {0.01~{\rm cm}} \Bigr ) ^2 + \Bigl( \frac{r_{\mathrm{trk}}-r_{\mathrm{PV}}}{\sigma_{r}} \Bigr)^2  \, .
    \end{equation}
    \item $ \atd$ : the ratio between the scalar $p_T$ sum of all tracks with $D_N < $ (certain value), normalized by the scalar $p_T$ sum of all tracks, hence $0 \leq \atd \leq 1 $.
\end{itemize}

\begin{figure}
    \centering
    \includegraphics[width=0.70\textwidth]{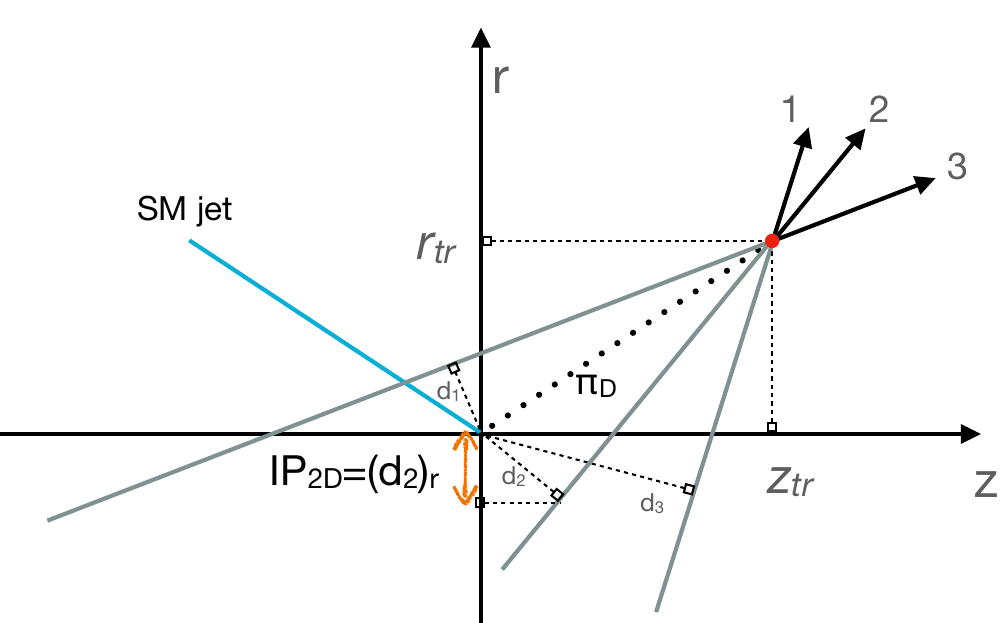}
    \caption{
Geometry of the considered variables. A SM jet and a dark pion $\pi_D$ originates from $(r,z)=(0,0)$; the latter decays at $(r_{tr}, z_{tr})$ into three tracks. We illustrate here how the $\iptd$ variable is obtained from the transverse impact parameter of the individual tracks.  See main text for details.
  }
    \label{fig:geometry}
\end{figure}

We illustrate the $\iptd$ variable in figure~\ref{fig:geometry}. From the primary vertex located at $(z,r) = (0,0)$ we consider that only a SM jet (which tags the vertex) and a long-lived dark pion emerge. The dark pion decays at $(z_{tr}, r_{tr})$ into three tracks $1,2,3$, which for illustration purposes we consider as giving rise to only one jet. The track trajectories meet at the decay vertex: the tracks are drawn in black, and their prolongations in grey. We indicate with $d_i$ the closest distance between the $i-$th track and the primary vertex, hence the $r$($z$) component gives the transverse (longitudinal) impact parameter. In the figure, the median of the $(d_i)_r$ corresponds to $(d_2)_r$,  hence the jet originating from the dark pion has $\iptd = |(d_2)_r|$.

Of the above variables, CMS presents results for $\iptd$  and $\atd$, before the selection cuts (which we describe below). Regarding the additional two variables, the variable $D_{N}$ should be small for tracks originating from prompt particles, and large for displaced tracks; while $\pudz$ is used for pile-up rejection. We note that both $D_N$ and $\pudz$ enter in these distributions through the definition of $\atd$, which depends on $D_N$. Since CMS has not made  explicit the $D_N$ threshold employed in their figure 3,  we have considered the three values employed to define the signal regions: 4, 10 and 20. Out of them, we have verified that the agreement is maximized for $D_N=10$, and hence $D_N < 10$ has been employed in the shown $\atd$ calculation shown below. As these last two variables are defined at the track level instead of at the jets level, we understand that kinematic distributions are not provided, which nonetheless would have provided an additional validation check for the proper reinterpretation of the search. 

Two important effects ought to be included for a realistic attempt at the reproduction of the CMS results: the tracking reconstruction efficiency $\epsilon_{\rm trk}$ and the smearing of the impact parameters.

CMS reports the tracking efficiency dependence in terms of the $p_T$, $\eta$, and transverse vertex position ($r$) of the track~\cite{CMS:2014pgm}. From figure 8 of this article, one can see that above $p_T > 1$ GeV one can consider the tracking efficiency independent of $p_T$ and to a lesser extent, of $\eta$. Regarding $r$, the efficiency diminishes with the displacement distance, as can be seen from figure 12 of~\cite{CMS:2014pgm}, which is obtained from a $t \bar{t}$ sample at $\sqrt{s}=7$ TeV. The figure shows the cumulative efficiency for each of the iterations (0-5) of the tracking algorithm. While this effect is less relevant for lifetimes of few millimeters, it has an impact for the benchmark point with $c \tau_{D} = 25$ mm and for larger lifetimes.

Nonetheless, it is clear that since the tracking efficiency is a very complicated function that can only be reliable obtained from having access to the full detector simulation (and detector information) we will pursue four different parametrizations for $\epsilon_{\rm trk}(r)$

\begin{itemize}
\item Use the reported value of Iteration 5 from figure 12 of~\cite{CMS:2014pgm}. [$It_5$]
\item Use the reported value of Iteration 4 from figure 12 of~\cite{CMS:2014pgm}. [$It_4$]
\item Consider that tracks with at least one hit in the inner detector are reconstructed with 100 \% efficiency, and with 0 \% if not: $\epsilon_{\rm trk}(r) = \Theta(r - 102~{\rm mm})$. [$R$].
\item Consider $\epsilon_{\rm trk}(r) = 1$, to illustrate the typical deviation obtained when no efficiency is considered. [$A$]
\end{itemize}

Regarding the impact parameter smearing, we note that for jets originating from SM quarks, one can expect to have $\iptd =0$, if the majority of the tracks of the light jet are prompt. However, the value of zero is obviously fictitious once the transverse impact parameter has been smeared to account for reconstruction effects. While the smearing functions have a non-trivial dependence with the $\eta$ and $p_T$ of the corresponding track (the resolution $\sigma_r$, which we have taken from figures 14a and 15a of~\cite{CMS:2014pgm}), the typical resolution would be of about 50 ${\rm \mu}$m, and hence the $\iptd$ variable would peak around this value for SM light-quark jets.

\begin{figure}[!htp]
    \centering
    \includegraphics[width=0.49\textwidth]{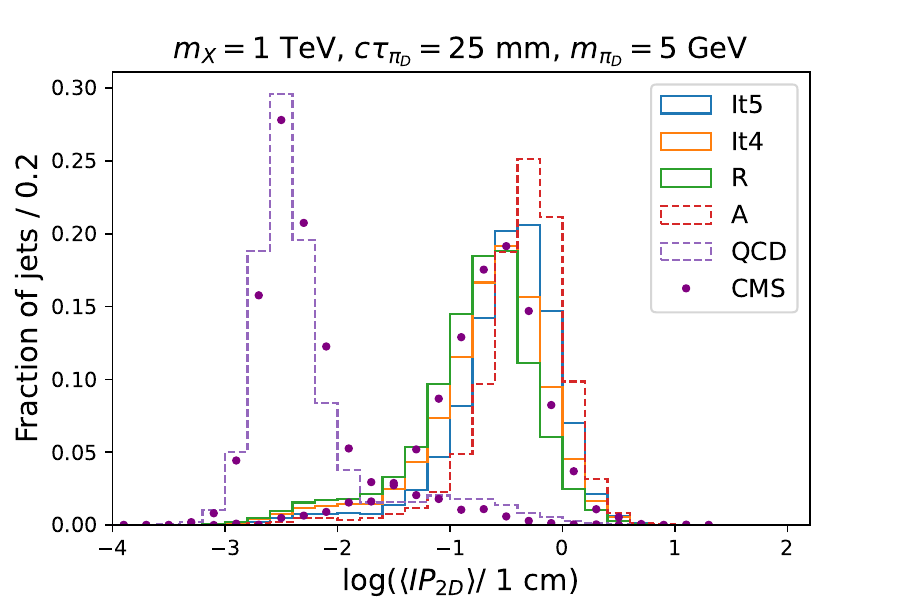}
        \includegraphics[width=0.49\textwidth]{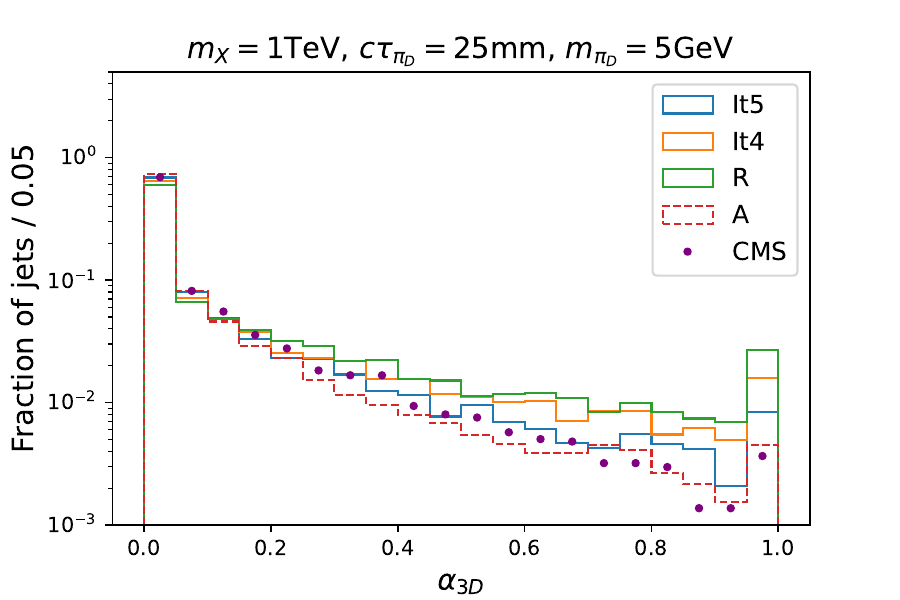}
    \caption{
    Comparison between the CMS published simulations and our Monte Carlo setup, for $\iptd $ (left) and $\atd$ (right) variables. For concreteness we only show the results for the benchmark point of $\mx= 1$ TeV, $\mpi= 5$ GeV and $\ctau = 25$ mm, yet a similar level of  agreement between the CMS data and our simulation was found for all benchmark points from~\cite{CMS:2018bvr}. We display results for the tracking efficiency parametrizations $It_5$ (solid blue), $It_4$ (solid orange), $R$ (solid green) and $A$ (dashed red). In the left panel, we include as well the results for the QCD jets coming from the $X$ decays in dashed purple.
  }
    \label{fig:validation}
\end{figure}

We show our results for the $\iptd$ and $\atd$ variables in Figure~\ref{fig:validation}, where we present the results for $A$ as dashed red, and for $It_5$, $It_4$ and $R$ in solid blue, orange and green. From the left panel we see that the naive $A$ approach does not describe the distribution as well as any of the other criteria, while the three other curves fit the signal distribution with reasonable accuracy. Moreover, we also see that the proper inclusion of the transverse impact parameter smearing is necessary to explain the distribution of $\iptd$ for the QCD jets from the signal, which is displayed as dashed purple.

From the right panel we see that the $\atd$ distribution for the signal does not change much with the different criteria $It_5$, $It_4$, $R$ and $A$. Since on this variable one only applies a $\atd < 0.25$ cut (see Appendix~\ref{a.CMS}) our attention is only in the proper reproduction of the first bins, and the mismatch at the tails is not relevant for us. Hence we delay the final judgement of which parametrization of the tracking efficiency to use to the next step of our validation: to reproduce the published $A_i$ efficiencies.
 
The analysis defined eight different jet identification criteria on the four relevant variables to consider a jet as \emph{emerging}. These criteria are supplemented by the requirement to have a minimum of two EJs , or one EJ jet with large transverse missing energy (MET), and by additional cuts on $H_T$ and on the $p_T$ of the four hardest jets. The combination of the EJ criteria and the additional cuts define seven \emph{selection sets}. The explicit requirements are collected in Appendix~\ref{a.CMS}, together with the 95 \% C.L. limit to the number of signal events in each selection set, $S_{95}^{i}$. 

\subsection{Reproducing efficiencies and exclusion limits}

\begin{figure}[!htb]
    \centering
    \includegraphics[width=0.455\textwidth]{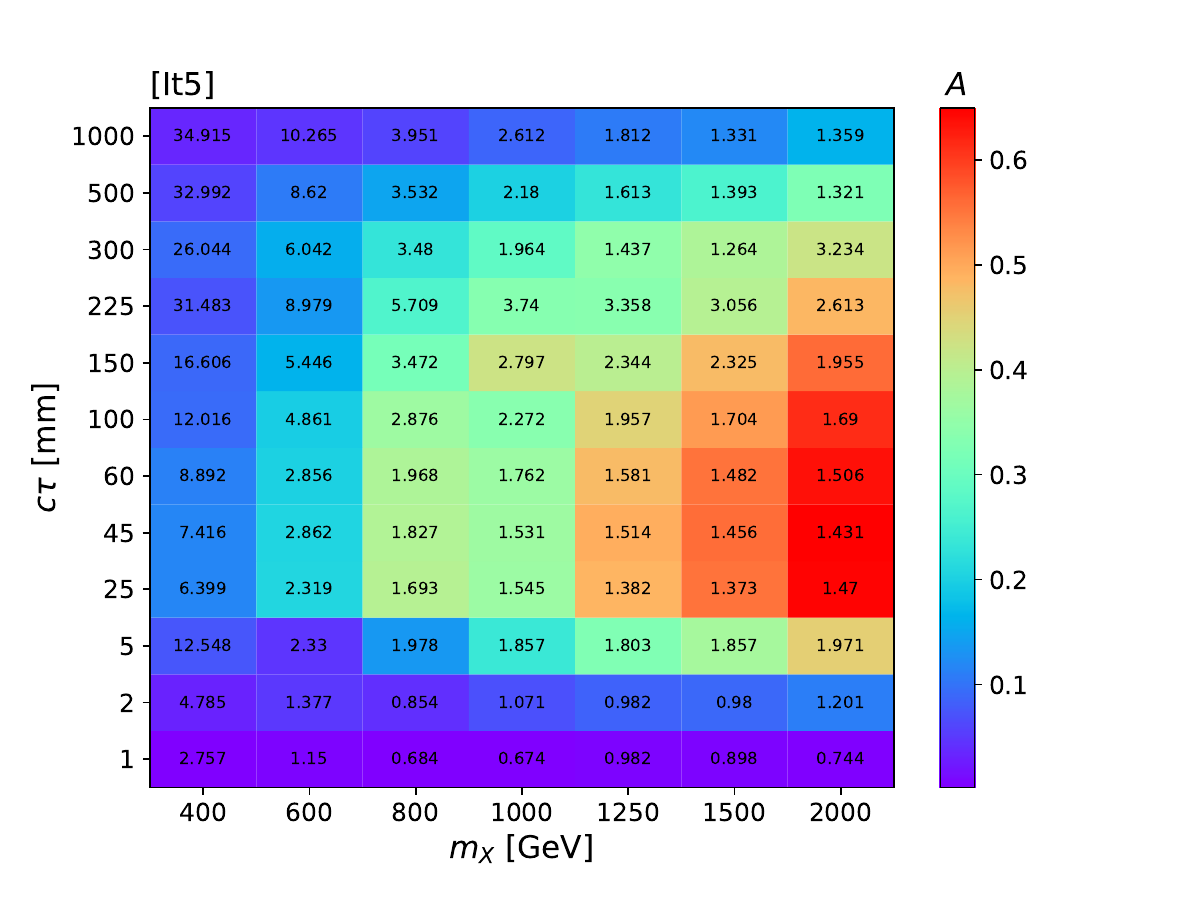}
    \includegraphics[width=0.455\textwidth]{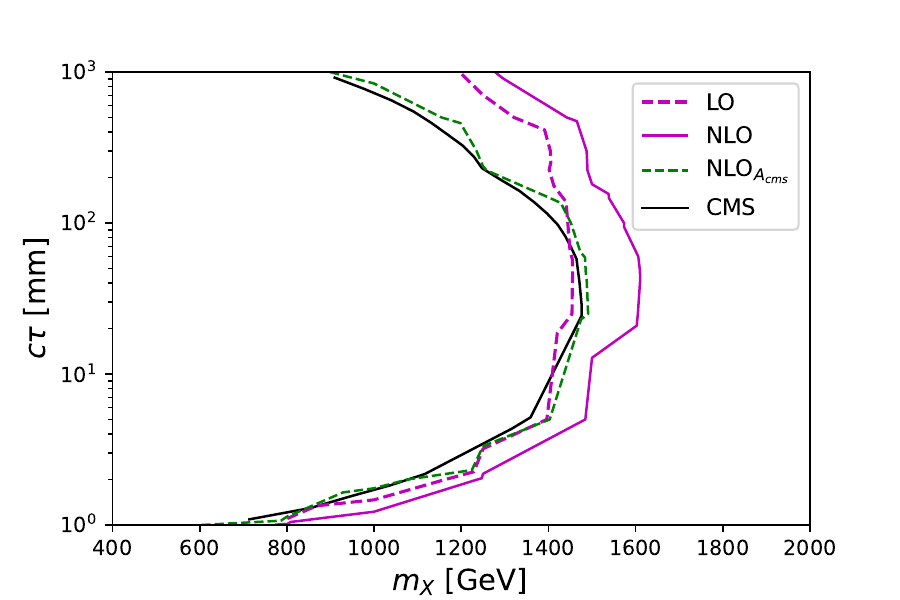}
    \includegraphics[width=0.455\textwidth]{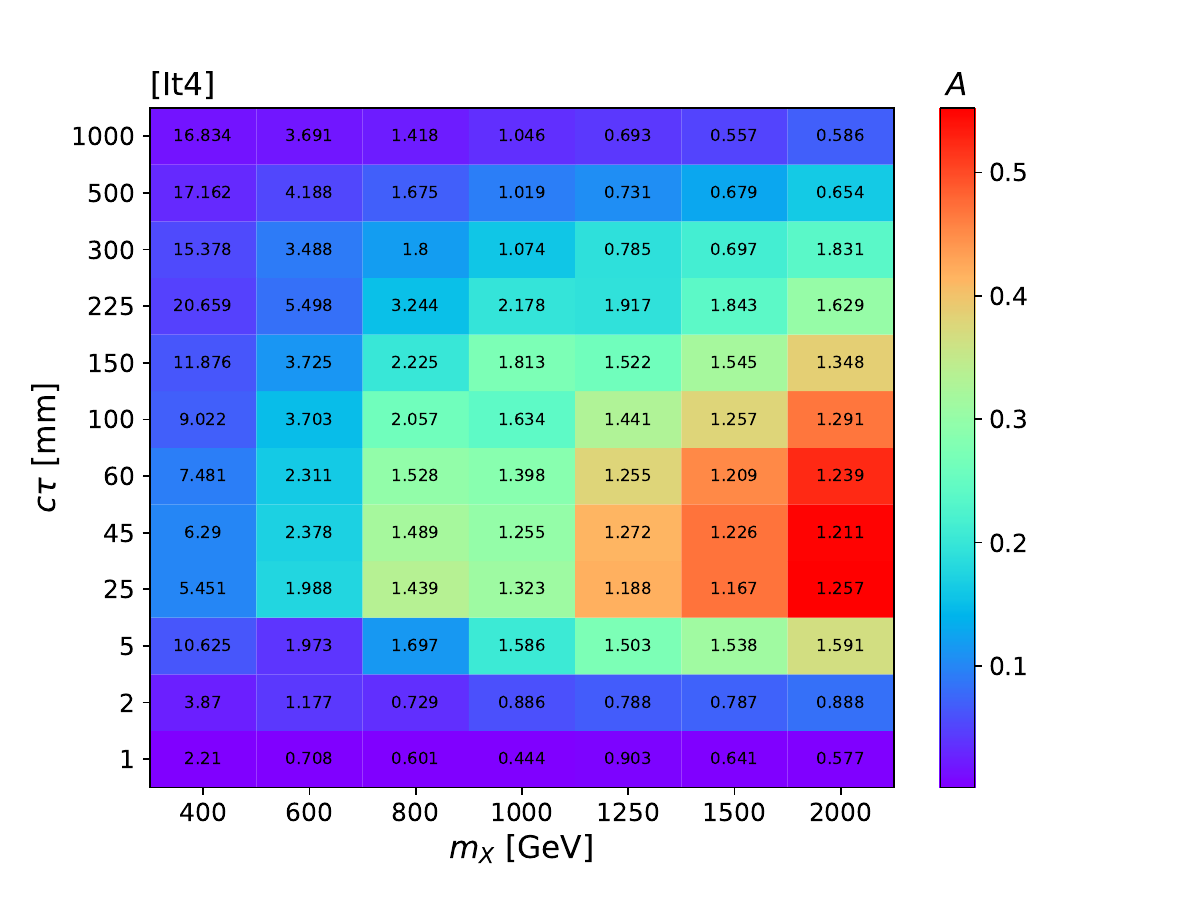}
    \includegraphics[width=0.455\textwidth]
    {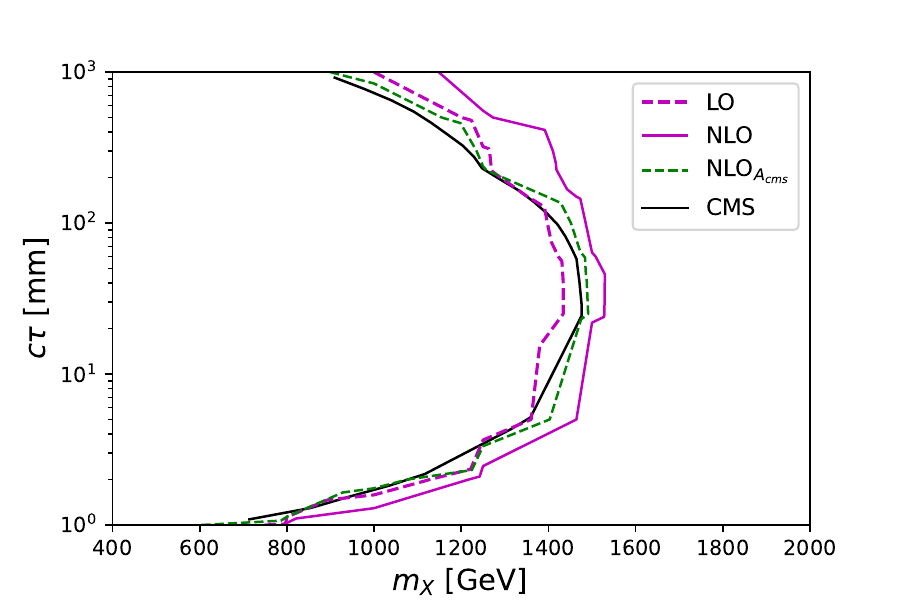}
    \includegraphics[width=0.455\textwidth]{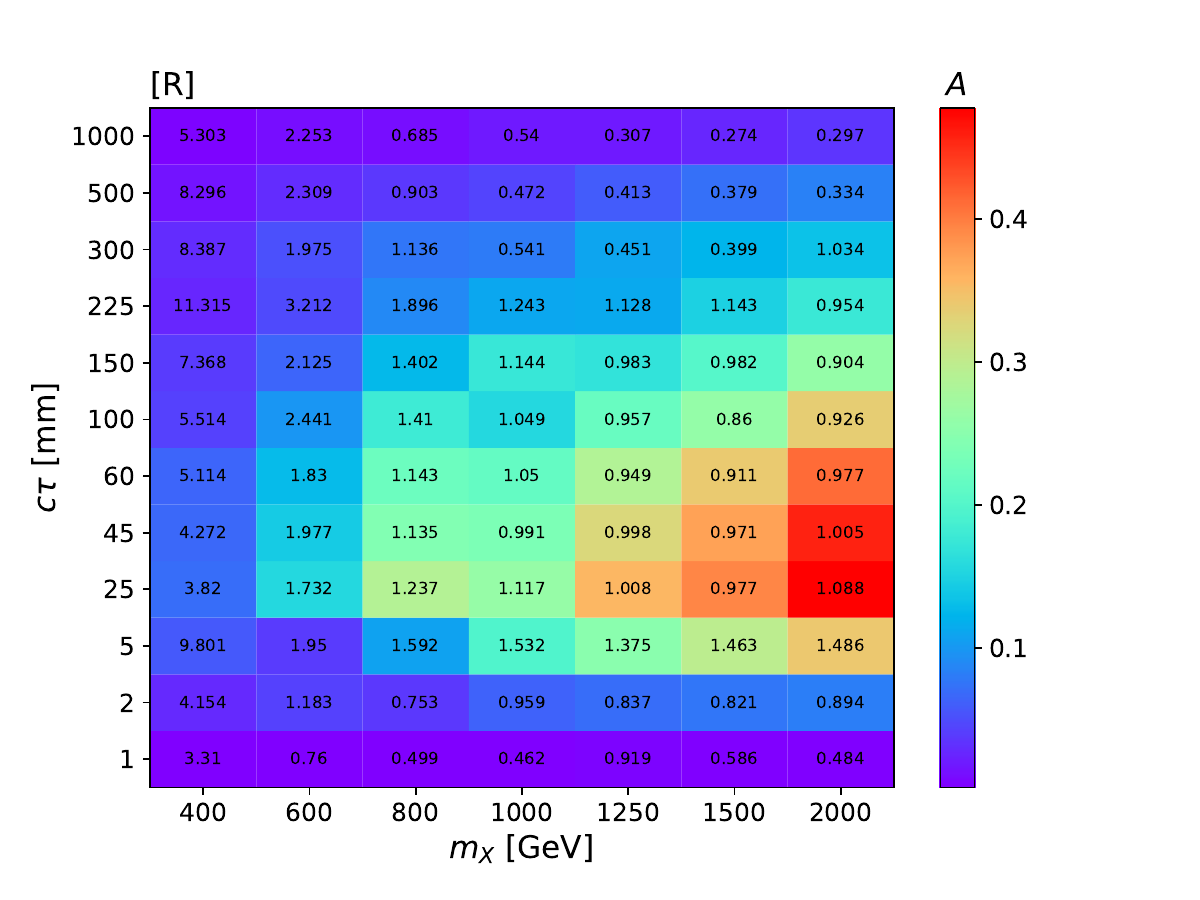}
    \includegraphics[width=0.455\textwidth]{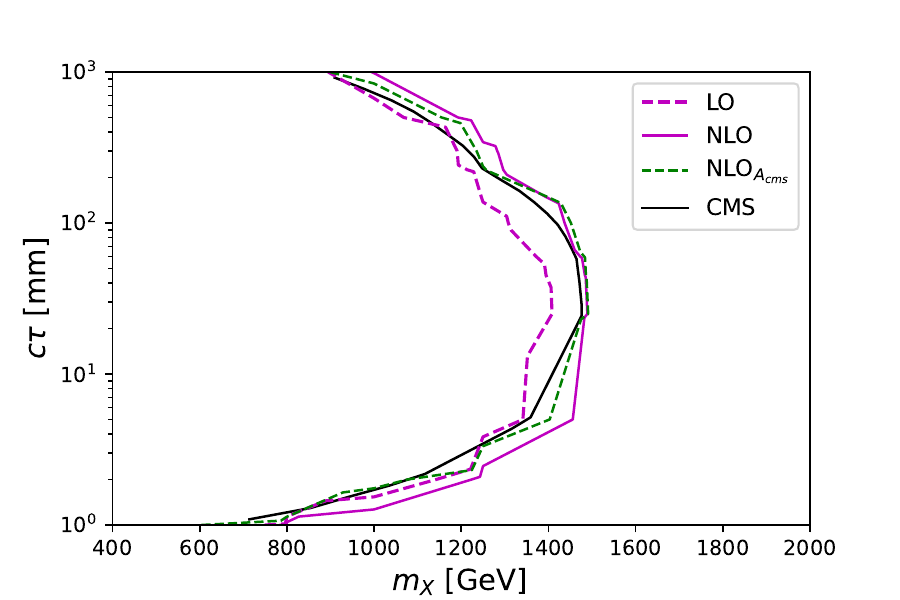}
    \caption{
    Ratio of our $A_i$ over those reported by CMS, for $m_{{\pi}_D} = 5$ GeV (left panels) and 95 \% exclusion limits (right panels) in the $m_X - c \tau$ plane. The tracking efficiency parametrization $It_5$, $It_4$ and $R$ have been used for the upper, middle and lower panels, respectively. In the left panel the colored bar indicates the value of $A_i$ and the displayed indicates the ratio of our $A_i$ normalized to the CMS one. In the right panel we show several exclusion limits: the CMS published one (solid black), the one obtained using $A_i$ CMS efficiencies with NLO predictions from figure~\ref{fig:limits} (dashed green), as well as those derived using our $A_i$ with LO (dashed pink) and NLO (solid pink) cross sections.  Both $It_4$ and $It_5$ tend to overestimate the acceptance (hence the exclusion limit), while the $R$ criteria reproduces better the exclusion curves. The disagreement is larger for the regions of lower masses and/or large lifetime, where the $A_i$ are below the per-mille level. }
    \label{fig:effvalidation}
\end{figure}

If our interest would be to perform a reinterpretation of the emerging jets results in the context of the \emph{same} model used by the collaboration (or one with a similar topology) then we could employ the reported acceptances $A_i$ to derive the published limits, as we did in Section~\ref{ss.pubAi}. We stress that our goal is to perform a flexible reinterpretation of this search, namely to employ it to derive limits on a model that the search has not considered. Hence, what we need is to fully validate our pipeline to compute the acceptance of the selection sets for the benchmark model used in the CMS study.  We show in Figure~\ref{fig:effvalidation} the ratio of our computed acceptances over the published CMS results in the left panels (the color bar indicate the $A_i$ value from CMS) and the obtained exclusion limits in the right panels, where we have employed the $It_5$ 
 (upper row), $It_4$ (middle row) and $R$ (lower row) parametrization of the tracking efficiencies. We can see that the best agreement is obtained with the $R$ parametrization, while the other two tend to overestimate the efficiencies. We see that we have agreement up to 20-30 \% for large masses in the iteration $R$, which degrades for lower masses and also extreme lifetime values, where the overall acceptances are nonetheless at the per-mille level or lower. We also note that the $R$ parametrization also gives an acceptable exclusion limit, and hence we decide to adopt it for the rest of the article. We note that with more examples provided by CMS (or simply by providing the efficiencies in all signal regions) one could attempt a more complex parametrization of the efficiency.

We consider hence this search as validated, and will proceed in the next section to derive bounds on the parameter space of Exotic Higgs decays. Our analysis code that allows us to derive the exclusions have been uploaded to the LLP Recasting Repository~\cite{LLPrepo}, making it publicly available to facilitate the reinterpretation of the emerging jets search for arbitrary models. Further instructions and the relevant documentation to run the code can be found in the Repository.

\section{Reinterpretation for Higgs mediated dark showers}
\label{s.results}

When the SM Higgs $h$ couples to the dark quarks the expected number of signal events reads
\be
N_S^{i} (\mpi, \ctau) = \sigma^{{\rm proc}}_{\rm SM} \times {\rm BR } (h \to Q_D Q_D)  \times A_i (m_H, \ctau, \mpi) \times {\cal L} \, ,
\ee
where now the only free physical parameters are the dark pion mass and its lifetime, and the exotic Higgs branching ratio into dark quarks. 

It is worth noting here that in Higgs resonant production the events would have a center of mass energy of the Higgs mass (approximately 125 GeV) while in the reinterpretation procedure from section~\ref{s.ejcms} the process of pair production of $X$ has a lowest mass of about 400 GeV. In the region of low masses our efficiencies overestimate the CMS result by a factor of a few. While the efficiency in those regions is quite low ( ${  \cal O}(10^{-3})$ and their uncertainty would depend on the Monte-Carlo statistics used in those benchmark points, it is also true that without any additional information in that region (e.g. kinematic distributions like those given for the $m_X = 1$ TeV benchmark) we can not investigate the origin of this discrepancy. Having expressed our reservations about the accuracy of our sensitivity estimates, we proceed with our analysis, taking the results \emph{cum grano salis}, and knowing that only a full-fledged experimental analysis can derive robust bounds. 

\begin{figure}[!htb]
    \centering
    \includegraphics[width=0.49\textwidth]{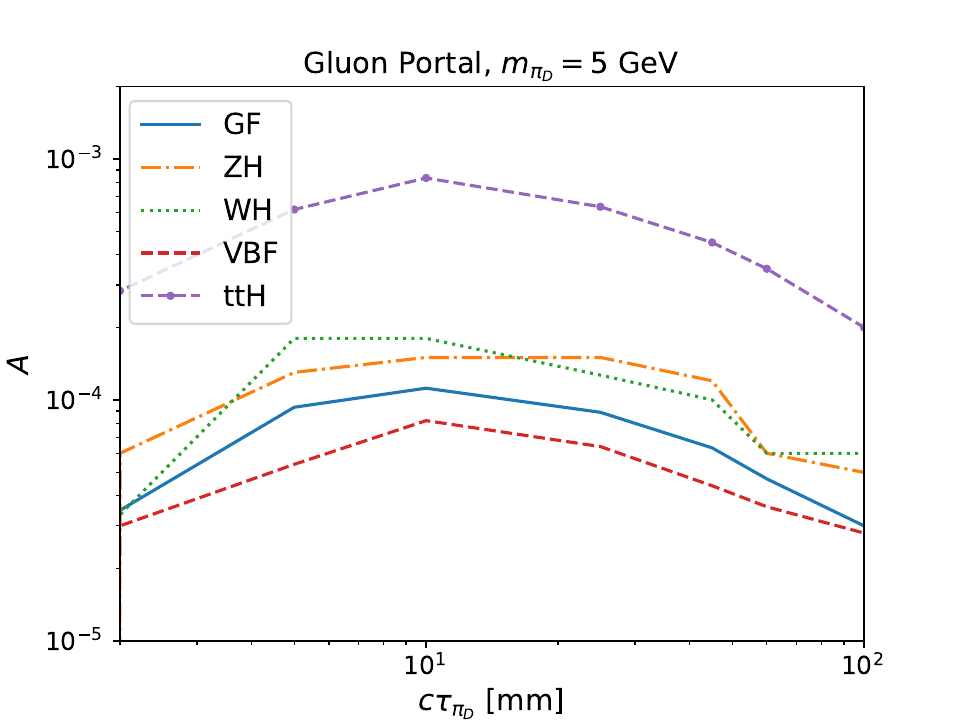}
    \includegraphics[width=0.49\textwidth]{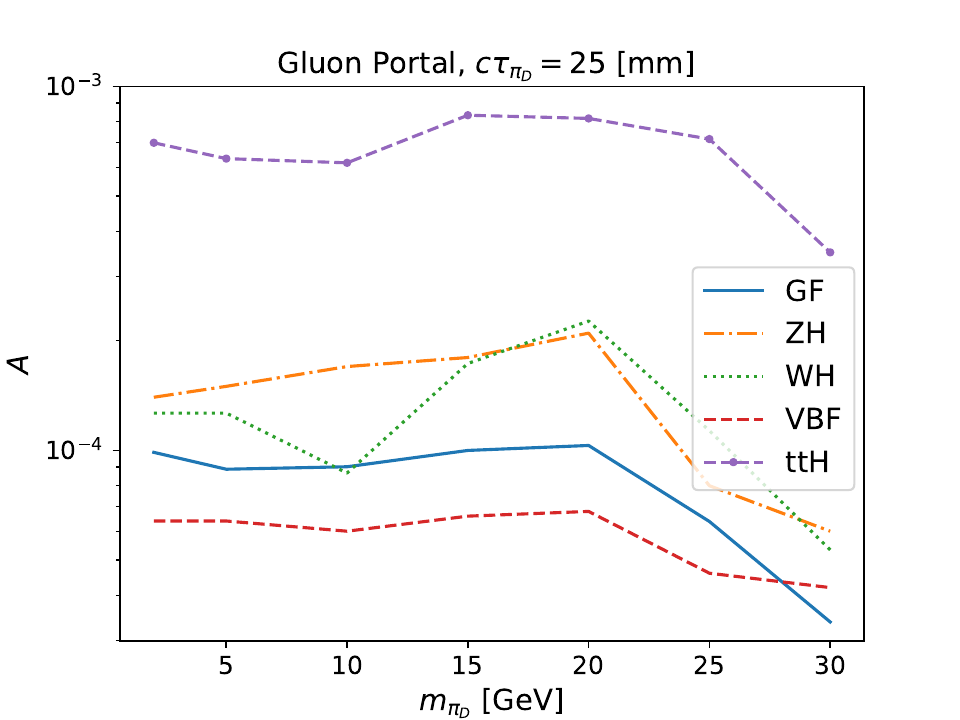}
    \caption{Maximum acceptance shown as a function of $\ctau$ (left) and $\mpi$ (right) in the Gluon Portal decay Benchmark for the Higgs production through Gluon Fusion (GF, solid), associated production with a Z,W boson (ZH: dot-dashed, WH: dotted), Vector Boson Fusion (VBF, dashed) and associated production with a $t \bar{t}$ pair (dashed with dots). } 
    \label{fig:acceptances}
\end{figure}

To further define our framework, we need to select a decay portal for our dark mesons. We follow here the proposal of reference~\cite{Knapen:2021eip} and we consider the gluon, vector, Higgs and dark-photon portals.\footnote{It is obvious that the photon portal, where $\pi_D \to \gamma \gamma$, does not pass the emerging jet selection cuts. Hence this decay portal is not further considered here.} We have verified our implementation of these decay portal benchmarks by reproducing the dark meson multiplicities from reference~\cite{Knapen:2021eip}.\footnote{We are indebted to Simon Knapen for useful communication during the validation phase.} We start by analyzing the acceptance $A_i$ as a function of the dark pion lifetime and masses, for the gluon decay portal, for all five considered production Higgs mechanisms, which we show in figure~\ref{fig:acceptances}. It is worth mentioning here that we do expect the EJ search not to be optimal for Higgs decays into dark quarks, as it originally targets two quarks and two dark quarks, while the Higgs decay would only give at parton level two dark quarks. However, since we are keeping the $\rho_D \to \pi_D \pi_D$ channel open, and since there is additional radiation from the initial state gluons and from the decay portals themselves, we do still obtain acceptances on the $10^{-4}$ range, which can suffice to obtain an exclusion given that with 16.1 fb$^{-1}$, ${\cal O} (10^{6} )$ Higgs bosons would be produced at the 13 TeV LHC via gluon-fusion. We illustrate the difference in kinematics between both models in Appendix~\ref{a.kinematics}, where we also explore in detail the impact of the event selection on the different Higgs production mechanisms.

From the figure we can see that the dependence in $\ctau$ is non trivial, obtaining a maximum around 10 mm, while for $\mpi$ the dependence is quite flat, except for the heavier masses of 20-30 GeV: those dark pions obtain a reduced boost from the Higgs compared to lighter ones.  It is intriguing to see that, owing to the additional radiation, the $ttH$ production has a higher acceptance, about an order of magnitude larger than gluon fusion, and about a factor of five larger than associated production with a vector boson. We note that vector fusion has the lowest acceptance, and this is due to the fact that the additional radiation in VBF goes in the forward direction, while the EJ analysis focuses on central jets. We stress that while we only show the gluon portal decay benchmark here, all the other portal decay models show an analogous behaviour. 

\begin{figure}[!htb]
    \centering
    \includegraphics[width=0.49\textwidth]{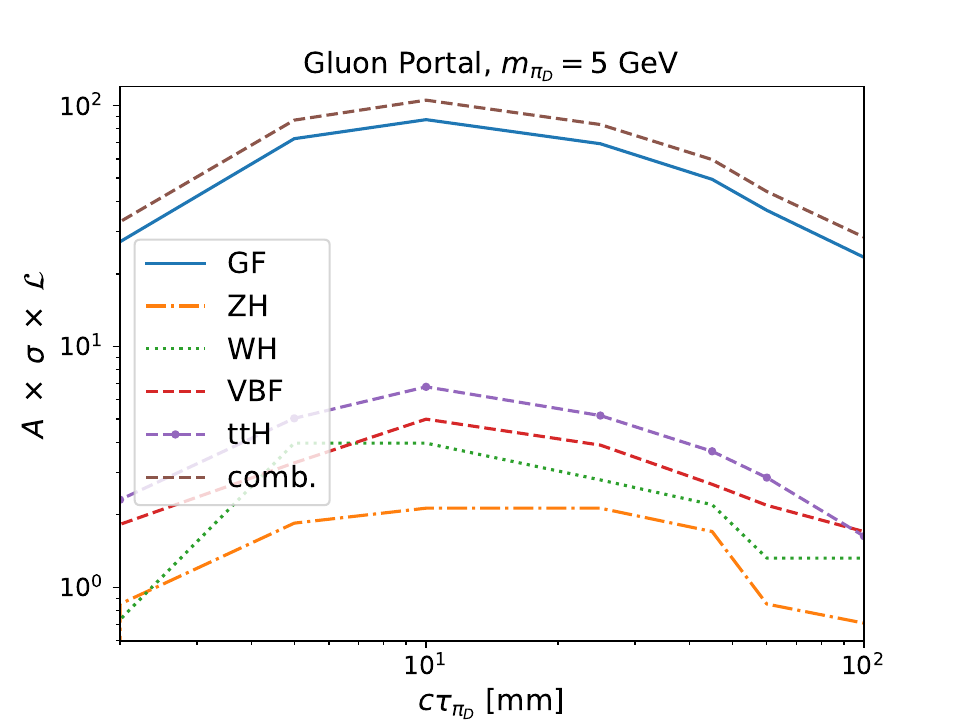}
    \includegraphics[width=0.49\textwidth]{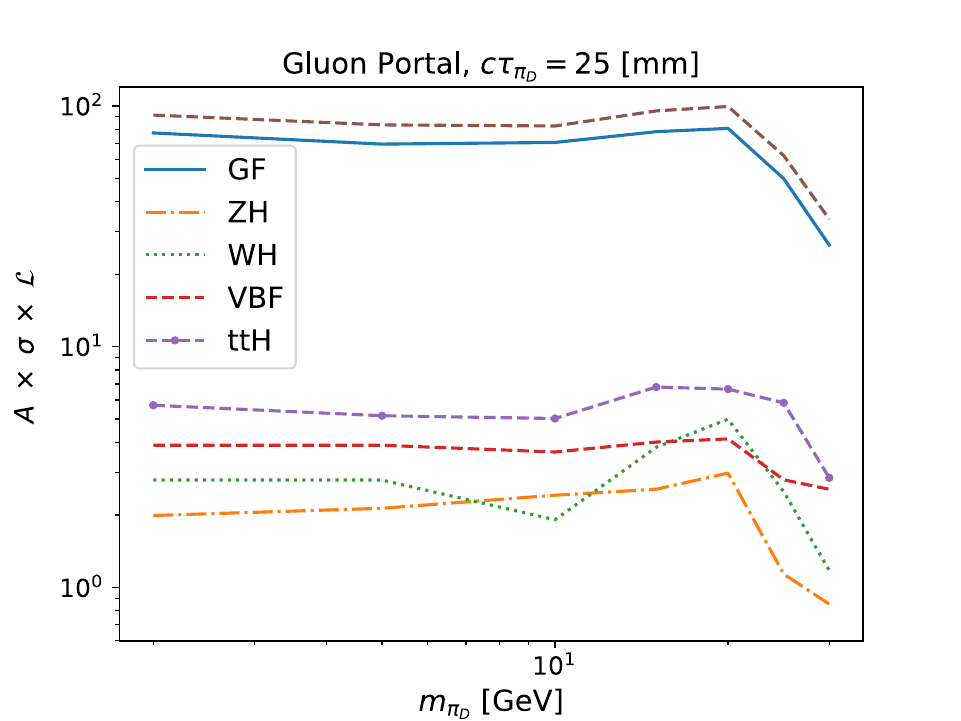}
    \caption{Maximum acceptance times production cross section times the total integrated luminosity of the EJ search (16.1 fb$^{-1}$) shown as a function of $\ctau$ (left) and $\mpi$ (right) in the Gluon Portal decay Benchmark for the Higgs production through Gluon Fusion (GF, solid), associated production with a Z,W boson (ZH: dot-dashed, WH: dotted), Vector Boson Fusion (VBF, dashed) and associated production with a $t \bar{t}$ pair (dashed with dots). } 
    \label{fig:xsacc}
\end{figure}

The picture changes slightly once the production cross sections for each mechanism is considered, which is shown in figure~\ref{fig:xsacc}. Here we multiply the maximum acceptance with the production cross section for each mechanism, and the total luminosity of the emerging jet search (16.1 fb$^{-1}$). Hence the y-axis directly displays the expected number of events for each production mode. We have added here the overall number of events obtained by summing over all possible production modes, in a dashed-brown line. We see now, that owing to the larger cross section of the GF mechanism (two orders of magnitude over $ttH$, factors of 15-25 for the modes involving gauge bosons), the overall number of events, $A \sigma {\cal L}$, is larger by an order of magnitude compared to the other modes. We also see that the impact of including all decay modes instead of only gluon fusion amounts to about 20 \% of the total number of events. In view of our findings we will focus from now on only on the dependence of our results with the lifetime for a $\mpi=5$ GeV mass, and we will also include all Higgs production modes in our study.

\begin{figure}[!htb]
    \centering
    \includegraphics[width=0.6\textwidth]{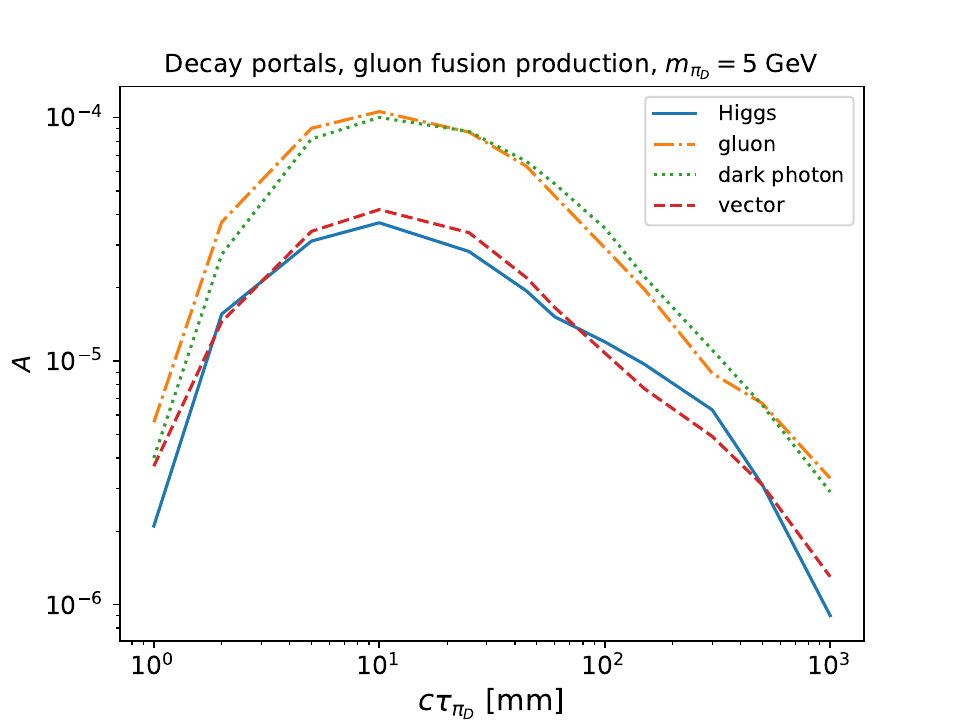}
    \caption{Acceptance for the different decay portals, for the gluon fusion production mode. We see that the gluon and dark photon portal have similar efficiencies, which are higher than those from the Higgs and vector portal, the latter pair also having similar efficiencies for the displayed mass. The acceptancres peak at a lifetime of about 10 mm. 
    } 
    \label{fig:eff_portals}
\end{figure}

We study now the sensitivity for the different decay portals considered in ref~\cite{Knapen:2021eip}. To that end we present in figure~\ref{fig:eff_portals} the efficiencies as a function of the dark pion lifetime, for $\mpi = $ 5 GeV. In order to obtain reliable estimates for these acceptances, we have simulated $10^{7}$ Monte Carlo events per parameter space point.

Of the possible decay portals, we then find that the sensitivity is larger (and similar) for the dark photon and gluon portals, and lower (and similar) for the vector and Higgs portals. Further details on the kinematics difference between the different portals can be found in Appendix~\ref{a.kinematics}. We then will select in what follows the gluon (G) and Higgs (H) decay portals, as they correspond to the extreme values for the efficiencies for the four portal scenarios considered.  These two decay portals correspond to the following operators
\be
\pi_D G^{\mu \nu} \tilde{G}_{\mu \nu} \qquad {\rm (G)}, \qquad \pi_D H^{\dagger} H \qquad {\rm (H)} \, .
\ee
In the gluon portal one expects a showered enriched with SM hadrons produced from the produced gluons, while in the Higgs portal the decays would follow a Yukawa-like structure, and one can expect a shower enriched with heavy flavour quarks.

\begin{figure}[!htb]
    \centering
    \includegraphics[width=0.75\textwidth]{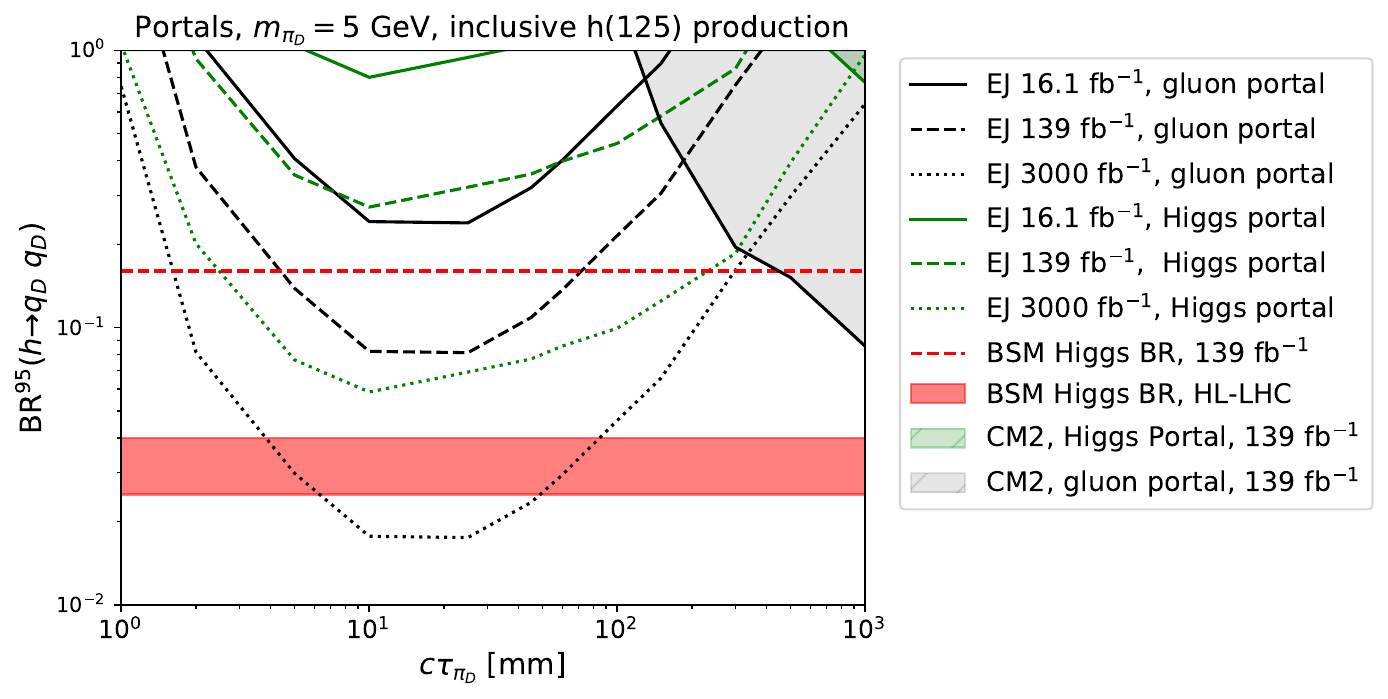}
    \caption{95 \% C.L limits on ${\rm BR} (h \to q_D q_D)$ obtained by reinterpreting the CMS emerging jet search, for the gluon portal and the Higgs portal as a function of $\ctau$ with $\mpi = 5$ GeV (left) and as a function of $\mpi$ with $\ctau = 25$ mm (right). The solid lines use the existing data, with a luminosity of 16.1 fb$^{-1}$, which correspond to the dataset of the current emerging jet search. The bounds from existing prompt searches, obtained with CheckMATE2~\cite{Dercks:2016npn} are shown as a hatched grey (green) region for the Higgs (gluon) portal. The dashed lines correspond to 139 fb $^{-1}$, the luminosity used on the current model-independent exclusions on undetected  (also called ``BSM'') Higgs branching ratios from ATLAS and CMS, while the dotted lines correspond to the projection to 3000 fb $^{-1}$, which we compare with the HL-LHC reach of the BSM Higgs branching ratios, shown with a red band that encapsulates the different assumptions on the systematic uncertainties (see main text for details).}
    \label{fig:EHDlimits}
\end{figure}

Using the acceptance from figure~\ref{fig:eff_portals}, we show in figure~\ref{fig:EHDlimits} the excluded exotic Higgs branching ratio as a function of the lifetime, for a dark pion mass of 5 GeV. The solid line is using the existing dataset from the EJ search (16.1 fb$^{-1}$). For comparison we show the ATLAS limit of 0.21, which was obtained with a 8 times larger dataset (139 fb$^{-1}$), shown in red dashed. For a fair comparison we rescale our EJ limit to this luminosity (dashed lines), assuming that the uncertainty is dominated by the statistical error, which given the event counts in the different signal regions, and the reported systematic errors, is a good approximation. We also include constraints from prompt searches using CheckMATE2 for prompt~\cite{Dercks:2016npn} and long-lived searches~\cite{Desai:2021jsa}, shown in dashed grey (green) for the gluon (Higgs) portal. These constraints come from prompt searches including missing energy from ATLAS, more concretely from~\cite{ATLAS:2019lng,ATLAS-CONF-2020-048}, using 139 fb$^{-1}$ of data: as the lifetime of the dark pions become large, many of them appear as missing transverse energy. We note that our benchmark choice for $\mpi = 5$ GeV correspond to a challenging phase space due to the softness of the decay products, otherwise it is clear that for heavier dark pion masses constraints from other searches would apply. One prominent example would be the 
 $Zh$ production, with $h$ decaying through light scalars or pseudoscalars into displaced jets~\cite{ATLAS:2021jig,CMS:2021yhb}, which report meaningful bounds on the exotic branching fraction for pseudoscalar masses as light as 15 GeV. Finally, we include an estimation of the HL-LHC, for both the BSM Higgs branching ratio (taken from~\cite{Cepeda:2019klc}) and for the statistic dominated approximation of the EJ search.  It is worth stressing that the HL-LHC run would not be similar to the current LHC setup in terms of capabilities to deal with long-lived particles (cite some HL-LHC expectations), significantly improving in many aspects, hence these limits must be considered with a grain of salt (also the statistical dominance of the study might not be fully justified).

From the figure we see that the EJ search can obtain better bounds than the model-independent BSM branching fraction, for lifetimes in the 5-60 (7-50) GeV when using the combined production (gluon fusion production) for the Higgs portal decay benchmark. For the sake of illustration we will describe now the bounds on this model, but the behaviour would be similar for other decay portals. Prompt searches with missing energy can constrain large lifetimes ($c \tau \gtrsim 400$ mm). For clarity reasons we have refrained from showing HL-LHC extrapolations from CheckMATE, but they would only be more sensitive than the BSM Higgs study for lifetimes in the 300-700 mm range, with the exact value depending on the final HL-LHC limit. Nonetheless, the phenomenological picture is similar to the one with the current dataset: for low lifetimes the BSM Higgs limit dominates, in an intermediate regime the EJ reinterpretation takes over, and for longer lifetimes the BSM Higgs searches become more sensitive, with missing energy searches becoming relevant for long-lifetimes.  As stressed before, exotic Higgs decays are not a target of the EJ analysis, and hence it would be interesting to consider the use of emerging jet taggers in other production modes. We leave this option for future work.

\section{Conclusions and Outlook}
\label{s.conclu}
In this work we have performed detailed studies focused on the reinterpretation of the CMS emerging jet search. This signature belongs to the class of signatures that are collectively dubbed as ``dark showers'', which stem from having a strongly-interacting dark (secluded) sector. In this dark sector new matter (and gauge) fields are added, which are assumed to hadronize, like in the SM strong sector. In particular, emerging jets correspond to the case where the dark sector mesons are have macroscopically appreciable decay lenghts, which make these final states also fall in the class of exotic phenomena dubbed ``long-lived particles'' (LLPs).

Our reinterpretation procedure has been validated by carefully following the CMS study. We have obtained good agreement with the published distributions on the $\iptd$ and $\atd$ variables, and also reproduced the publicly available efficiencies for the benchmark model employed in the search. We have reproduced the published exclusion limits through two different routes, one by employing directly the CMS published efficiencies and another one by computing the efficiencies ourselves through our own Monte-Carlo simulation. Here there is a large uncertainty in the exact parametrization of the tracking efficiency. We have attempted a few different parametrizations, and employed the one that, while possibly over-simplified, can reproduce the published efficiencies (and exclusion limits) with a reasonable accuracy.

We would like to stress that while the relevant information of the CMS study was publicly available and clearly explained, getting in contact with the authors of the experimental study was nonetheless needed in order to comprehend a few crucial details. Their response has been instrumental to understand details concerning the track efficiency and the impact parameter smearing used in the study.  Since it would be desirable that a reinterpretation of an experimental study can be done without this contact (as it can happen that the main authors of a given analysis might not be always part of the collaboration), we also took the opportunity to comment in the text for which aspects a clarification was needed, and which additional material would have helped us to carry our the reinterpretation. 

Using our validated pipeline, we have focused on the exploration of a SM Higgs boson decaying into two dark quarks (fermions charged solely under the new strong sector, akin to the SM quarks). To that extent, we have considered the inclusive production of the Standard Model Higgs from gluon fusion, Higgs-strahlung, vector-boson-fusion and associated production with a $t \bar{t}$ pair, and analyzed four decay benchmark portal models proposed in~\cite{Knapen:2021eip}, which are dubbed gluon, dark photon, Higgs and vector portals. We have found that, while the efficiencies for the Higgs production rank in the 10$^{-3:-5}$ range, owing to the large production cross section we can obtain meaningful bounds in the relevant parameter space, which are competitive with the current exclusion on \emph{undetected} Higgs branching ratio of 16 \%, set by the ATLAS and CMS collaborations. We have checked, with the help of CheckMATE, that the existing prompt searches can bring meaningful bounds only for the large lifetime regime, $\ctau \gtrsim {\cal O} (100 {\rm mm})$. We have also considered the existing HL-LHC extrapolations for the \emph{undetected} Higgs branching ratio, and compared them with a similar naive extrapolation of the emerging jet search sensitivity (relying only on statistical uncertainties being present). Yet, it is expected that the HL-LHC will have a number of improvements to detect long-lived particles, which could render the final projections better than our naive extrapolations.

As a byproduct of our analysis, we have made publicly available our Pythia 8 analysis code in the LLP Recating Repository~\cite{LLPrepo}, which can be used to compute the experimental acceptance (and the exclusion limits) with arbitrary BSM models, provided they are implemented in Pythia8.

We would like to stress that the exotic Higgs decay exclusion from~\cite{ATLAS:2019nkf} is an indirect bound, based on a global fit to the observed Higgs properties. Hence, if a signal is detected, its characterization would require an independent study. In contrast, if the emerging jet search starts seeing an excess, one can already infer that a new long-lived object is being produced from a Higgs boson decay, information that is crucial for the proper characterization of a putative BSM signal.

We end by noting that the EJ requirements of having four hard jets do not precisely target the exotic decays of a SM Higgs boson. In spite of the analysis not being optimal, we see that we can exclude exotic branching ratio of 30 \% in the gluon and dark photon decay portals, which can go down to the percent level for HL-LHC. Therefore, it might be worthwhile to explore EJ searches that focus on dark quark decays from a SM Higgs boson (or from a new scalar), which could have higher sensitivity than the model independent search for undetected Higgs branching ratios.

\subsection*{Acknowledgements}
We would like to thank Juliette Alimena, Nishita Desai, Alberto Escalante del Valle, Simon Knapen, Emmanuel Francois Perez  and Pedro Schawaller for useful discussions, and Baibhab Pattnaik for a careful reading of the manuscript. We are indebted to the authors of the CMS emerging jet analysis: Alberto Belloni, Yi-Mu Chen, Sarah Eno and Long Wang for their patience to answer our questions about technical details in their study. 
JC and JZ are supported by the {\it Generalitat Valenciana} (Spain) through the {\it plan GenT} program (CIDEGENT/2019/068), by the Spanish Government (Agencia Estatal de
Investigación) and ERDF funds from European Commission (MCIN/AEI/10.13039/501100011033, Grant No. PID2020-114473GB-I00). JC is also supported by the {\it Generalitat Valenciana} (Spain) through the {\it plan GenT} program (CIDEGENT/2019/068) and (CIDEGENT/2018/014).

\appendix

\section{CMS emerging jets}
\label{a.CMS}
In this Appendix we include additional material from our CMS validation described in ~\ref{s.ejcms}.

The CMS collaboration employs four variables to identify emerging jets in both signal and background regions, which have been defined in section~\ref{ss.kin}. For the six signal regions, the specific requirements on each of these variables are shown in Table~\ref{t.ejrequirements}.

\begin{table}[!htp]
\begin{tabular}{ccccc}
\hline
Group & $\pudz$ > ... cm & $D_\mathrm{N}$ < ... & $\iptd$ > ... cm & $\atd$ < ... \\ \hline
EMJ-1                               & 2.5                                      & 4                            & 0.05                                     & 0.25                                             \\
EMJ-2                               & 4.0                                      & 4                            & 0.10                                     & 0.25                                             \\
EMJ-3                               & 4.0                                      & 20                           & 0.25                                     & 0.25                                             \\
EMJ-4                               & 2.5                                      & 4                            & 0.10                                     & 0.25                                             \\
EMJ-5                               & 2.5                                      & 20                           & 0.05                                     & 0.25                                             \\
EMJ-6                               & 2.5                                      & 10                           & 0.05                                     & 0.25                                            
\end{tabular}
\caption{CMS requirements to tag a given jet as ``emerging''. The variables are defined in the main text.}
\label{t.ejrequirements}
\end{table}

Based on these requirements, CMS further defines signal regions (called ``sets'' in the CMS paper), where a given EMJ criteria is accompanied by a set of cuts on the jets, requiring either two emerging jets, or one emerging jet plus large missing transverse energy. Those definitions are shown in Table~\ref{t.SR}. The event yield excluded in each signal region at the 95 \% C.L. by CMS is shown in the rightmost column, $S_{95}$.

\begin{table}[!htb]
\begin{tabular}{ccccccccccc}
\hline
Set \# & $H_T$ & $p_T(1)$ & $p_T(2)$ & $p_T(3)$ & $p_T(4)$ & $p_T^{\rm miss}$ & $n_{\mathrm EMJ} >= $ & EMJ group & $S_{95}$ \\ \hline
1                           & 900   & 225      & 100      & 100      & 100      & 0                & 2                        & 1         & 36.7                        \\
2                           & 900   & 225      & 100      & 100      & 100      & 0                & 2                        & 2         & 14.6                        \\
3                           & 900   & 225      & 100      & 100      & 100      & 200              & 1                        & 3         & 15.6                        \\
4                           & 1100  & 275      & 250      & 150      & 150      & 0                & 2                        & 1         & 15.1                        \\
5                           & 1000  & 250      & 150      & 100      & 100      & 0                & 2                        & 4         & 35.3                        \\
6                           & 1000  & 250      & 150      & 100      & 100      & 0                & 2                        & 5         & 20.7                        \\
7                           & 1200  & 300      & 250      & 200      & 150      & 0                & 2                        & 6         & 5.61                       
\end{tabular}
\caption{Signal regions defined in the Emerging Jet study by CMS. The $S_{95}$ column indicates the 95 \% C.L on the expected amount of signal events, which we employ for limit setting.}
\label{t.SR}
\end{table}

The information from these tables has been included in the companion code uploaded to the LLP Recasting Repository~\cite{LLPrepo}. We have also collected there the details on the different tracking efficiency parametrization employed in this work.

\section{Emerging jets kinematics}
\label{a.kinematics}
In this Appendix we provide further details on the kinematic differences between the benchmark model proposed by Schwaller, Stolarski and Weiler~\cite{Schwaller:2015gea} (SSW) used by CMS, and the Higgs-mediated dark showers we employed in our study.

\begin{figure}[!htb]
    \centering
    \includegraphics[width=0.455\textwidth]{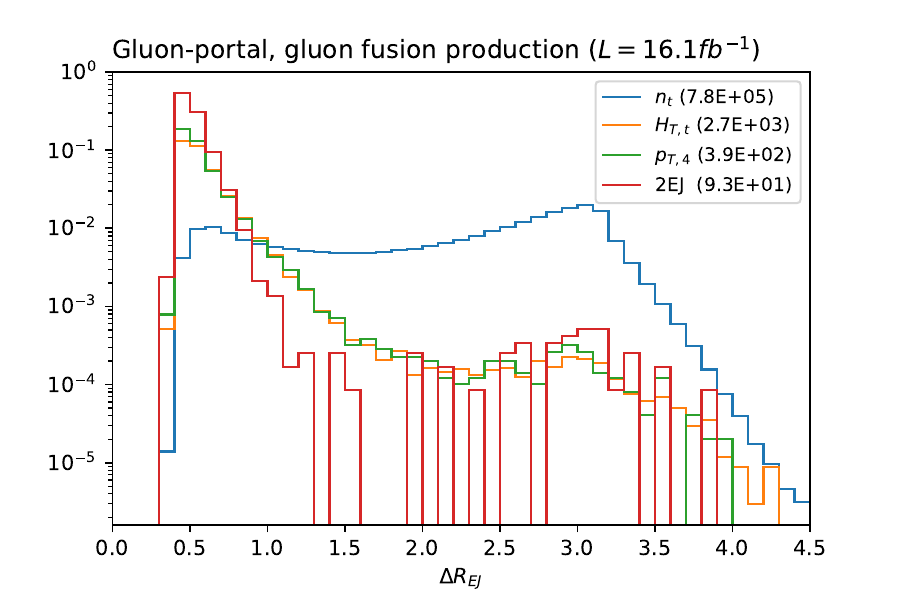}
    \includegraphics[width=0.455\textwidth]{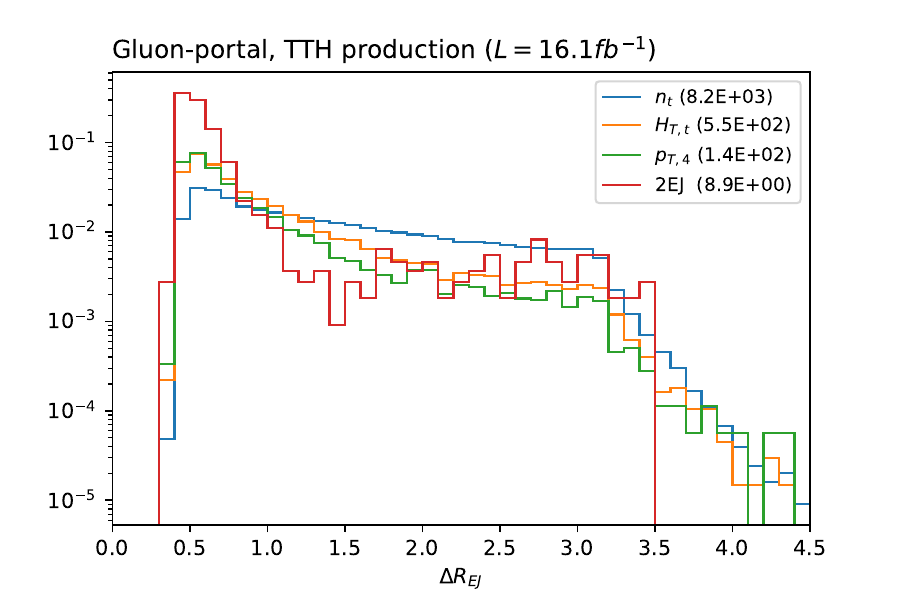}
    \includegraphics[width=0.455\textwidth]{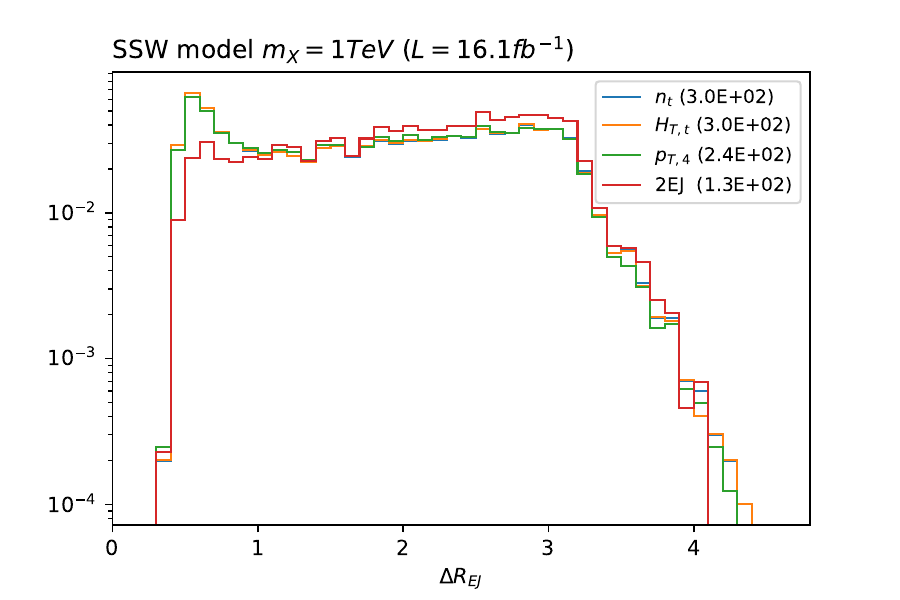}
    \includegraphics[width=0.455\textwidth]{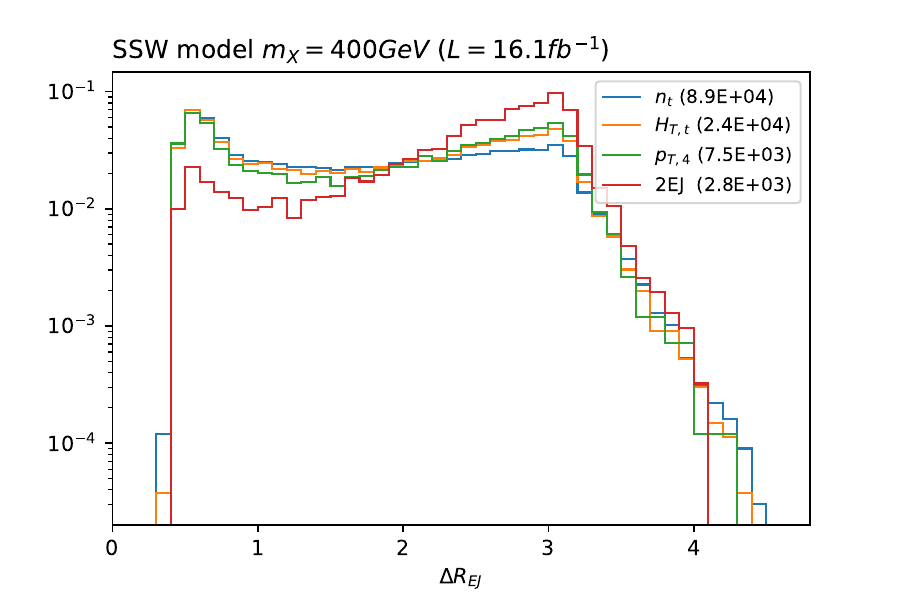}
    \caption{$\Delta R$ between the hardest two emerging jets after applying different selection cuts (consecutively). We present results for the gluon portal decay mode, for gluon fusion production (upper left) and $ttH$ production (upper right); and for the SSW model for the $m_X=1$ TeV (lower left) and $m_X=400$ GeV (lower right). We also show the expected number of events, with $\emph{L}=16.1~{\rm fb}^{-1}$, in parenthesis.}
    \label{fig:deltaREJs}
\end{figure}

We start by showing the angular distance $\Delta_R$ between the two hardest emerging jets in figure~\ref{fig:deltaREJs}. Here we fixed the lifetime of the dark pion at 10 mm, which is the value at which the efficiencies peak, and the emerging jets are reconstructed here using the requirements EMJ-1, see Table~\ref{t.ejrequirements}. We show these distributions at four different steps of the event selection. The blue curve show all events that come out of the Monte Carlo simulation, without imposing any cuts, $n_T$, where we request the presence of two emerging jets. The orange curve show the events that pass the $H_T > 900$ GeV trigger, $H_{T,t}$. The green curve shows those events where the $p_T$ conditions on each jet are applied, using Set \# 1 from Table~\ref{t.SR}. Finally, in the red curve we request that the two tagged emerging jets are included in the set of the four hardest jets . For clarity reasons we have normalized all the distributions to unity, but in the legend we show in parentheses the number of expected signal events for the luminosity of the CMS study, 16.1 fb$^{-1}$. We consider here the decay portal with the largest efficiency (gluon portal) and compare gluon fusion production (upper left) with associated Higgs production with a $t \bar{t}$ pair (upper right). For comparison purposes we also show results for the SSW model used by CMS, in their $m_X = 1$ TeV benchmark, and show the impact for lower masses, taking $m_X= 400$ GeV (lower right).

\begin{figure}[!htb]
    \centering
    \includegraphics[width=0.6\textwidth]{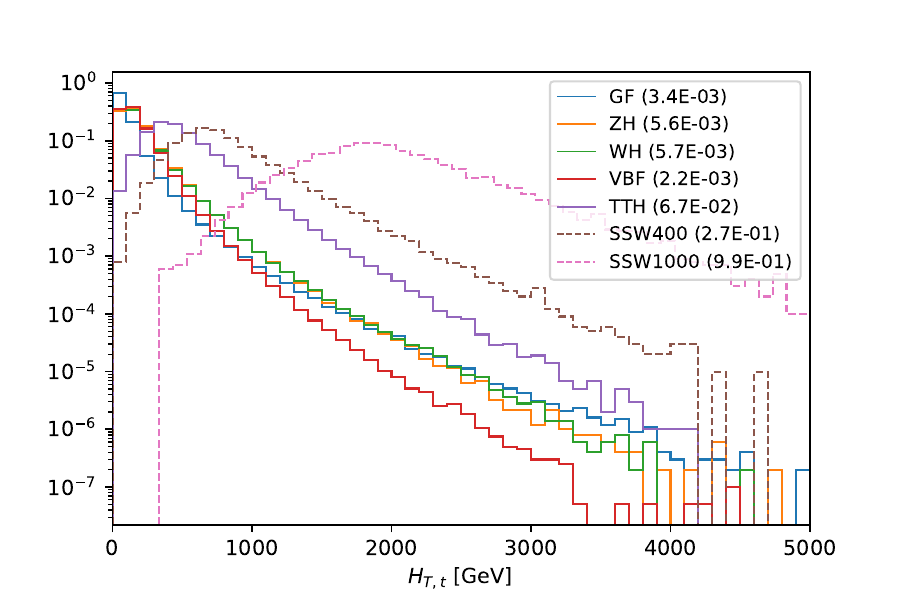}
    \caption{$H_T$ distributions (normalized to unity) before applying selection cuts and the trigger condition. We show for the gluon portal, the five production mechanisms considered, and we also show the SSW benchmarks with 400 and 100 GeV mediator masses. The number in parenthesis in the legend indicates the efficiency of the $H_T > 900 $ GeV trigger selection. }
        \label{fig:HTprod}
\end{figure}

From the figure we see that indeed the large boost of the Higgs when the $H_T$ trigger condition is applied forces the EJ to be highly boosted, as now the distribution falls very fast with $\Delta R$. This suggests that the use of jet-substructure techniques (see e.g.~\cite{Salam:2010nqg}), which have already been applied to semi-visible jets in~\cite{Cohen:2020afv}, could be of great help to increase the sensitivity for emerging jets. We also see from the numbers within parenthesis that the overall acceptance is driven by the large $H_T$ requirement. For gluon fusion the $H_T$ cut has an efficiency of $3.5 \times 10^{-3}$, out of the overall $1.24 \times 10^{-4}$ efficiency. For $ttH$, we see that the $H_T$ cut has an efficiency of $6.7 \times 10^{-2}$, out of the overall $1 \times 10^{-3}$ efficiency. We note that this behaviour happens as well for the other production modes not shown here (VBF, WH, ZH). We can see the impact of the $H_T$ trigger on the distributions shown in figure~\ref{fig:HTprod}. Indeed, the $H_T$ shape that clearly distinguishes $ttH$ from the other production mechanism (and also between different mediator masses in the SSW model) explains the outcome of our acceptance plots.

It is also important to note that the impact of $H_T$ also drives the efficiency for the different decay portals. In table~\ref{tab:cutflow} we show the cutflow for the higgs, gluon, darkphonton and vector portal, in gluon fusion production. Once again, we see that the $H_T$ trigger requirement is what drives the overall efficiency. This provides an important motivation to consider other triggers for Higgs-mediated dark showers.

\begin{table}[!htb]
    \centering
    \begin{tabular}{c|c c c c}
        Cuts & gluon & Higgs & dark photon & vector \\ \hline
        $H_T > 900$GeV &  0.00342 & 0.00238 & 0.00343 & 0.00217 \\
        $p_{T,j1}>225$GeV & 0.00209 & 0.0014 & 0.00209 & 0.00126 \\
        $p_{T,j2}>100$GeV & 0.00209 & 0.0014 & 0.00209 & 0.00126 \\
        $p_{T,j3}>100$GeV & 0.00145 & 0.00098 & 0.00146 & 0.00089 \\
        $p_{T,j4}>100$GeV & 0.00049 & 0.00032 & 0.000494 & 0.0003 \\
        $n_{EJ} \geq 2$ & 0.00012 & 0.00004 & 0.00012 & 0.00005 \\
    \end{tabular}
    \caption{Cutflow of the signal efficiency for the different decay portals: gluon, Higgs, dark photon and vector portal, considering gluon fusion production for the benchmark point $m_{\pi_D}=5$ GeV and $c\tau_{\pi_D}=10$ mm.}
    \label{tab:cutflow}
\end{table}

\begin{figure}[!htb]
    \centering
    \includegraphics[width=0.455\textwidth]{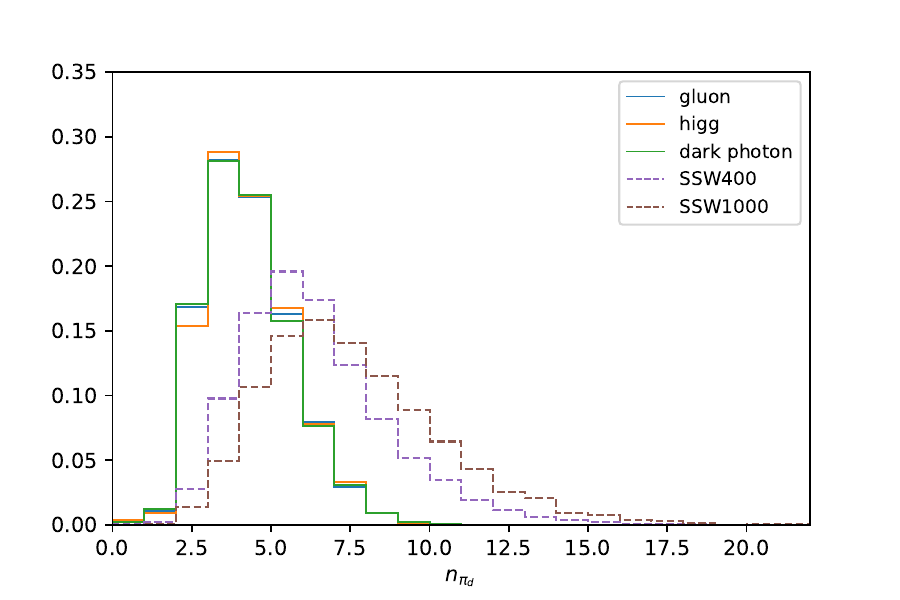}
     \includegraphics[width=0.455\textwidth]{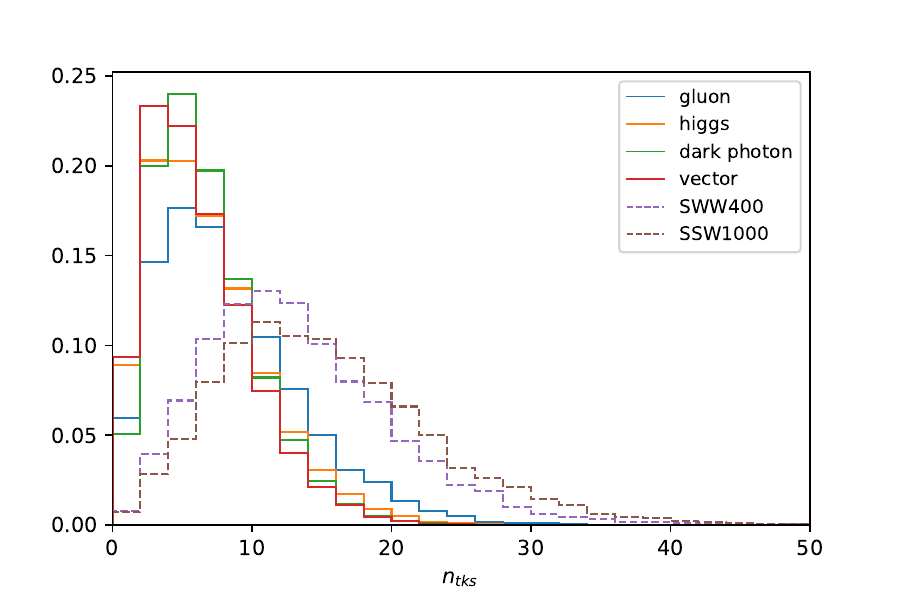}
    \caption{Dark pion (left) and track multiplicity (right) for Higgs production through gluon fusion in the gluon (blue), higgs (orange), dark photon (green) and vector (red) portals; and also for the SSW model with $m_X = 400$ (dashed purple) and $m_X=1$ TeV (dashed brown). }
    \label{fig:multiplicity}
\end{figure}

Finally, we would like to provide some insight on the kinematic of the different decay portals. We show the dark pion multiplicity, and the track (within each emerging jet) multiplicity in figure~\ref{fig:multiplicity}, fixing the production mode to gluon fusion, and considering the gluon (blue), higgs (orange), dark photon (green) and vector (red) portals. In addition we also include for comparison purposes the SSW model with the 400 GeV and 1 TeV mediator masses. From the left panel we see that the overall shower multiplicity is governed mostly by the Higgs mass, irrespective of the decay portal. This can be seen when comparing the two SSW benchmarks, how the multiplicity decreases with the mediator mass. The difference between the portals, however, appears when considering the track multiplicity within each emerging jet. Since the track kinematics is being used to actively tag the jets, these differences explain why the gluon and dark photon portals have larger acceptance than the Higgs and vector portals. 

\bibliographystyle{JHEP} 
\bibliography{EJreint}

\end{document}